# Mapping finite-fault slip with spatial correlation between seismicity and point-source Coulomb failure stress change

Anthony Lomax*[1]

[1]ALomax Scientific, Mouans-Sartoux, France
*Corresponding author: anthony@alomax.net

**Author ORCIDs**

A. Lomax: 0000-0002-7747-5990

**Abstract**

Most earthquake energy release arises during fault slip many kilometers below the Earth's surface. Understanding earthquakes and their hazard requires mapping the geometry and distribution of this slip. Such *finite-fault* maps are typically derived from surface phenomena, such as seismic and geodetic ground motions. Here we introduce an imaging procedure for mapping finite-fault slip directly from seismicity and aftershocks—phenomena occurring at depth around an earthquake rupture. For specified source and receiver faults, we map source-fault slip in 3D by correlation of point-source Coulomb failure stress change (ΔCFS) kernels across the distribution of seismicity around an earthquake. These *seismicity-stress* maps show relative, static fault slip compatible with the surrounding seismicity given the physics of ΔCFS; they can aid other slip inversions and aftershock forecasting, and be obtained for early instrumental earthquakes. We verify this procedure recovers synthetic fault slip, and matches independent estimates of slip for the 2004 Mw 6.0 Parkfield and 2021 Mw 6.0 Antelope Valley California earthquakes. For the 2018 Mw 7.1 Anchorage Alaska intra-slab earthquake, seismicity-stress maps, combined with multi-scale precise hypocenter relocation, resolve the enigma of which mainshock faulting plane ruptured (the gently east-dipping plane), and clarify slab structures activated in the energetic aftershock sequence.

**Non-technical summary**

The energy and shaking from earthquakes radiates from slip across faults many kilometers deep in the Earth. Mapping the geometry and strength of this fault slip for large earthquakes is fundamental for understanding earthquakes and their hazard. Usually such mapping is done with observations at the Earth's surface. Here we introduce a slip mapping procedure using aftershocks, which form virtual observations surrounding the earthquake at depth. A large earthquake produces extensive deformation and consequently aftershocks in the surrounding rock. The distribution of aftershocks is irregular but can be estimated from the orientation of the large earthquake fault and an assumed orientation for the aftershock faults. By comparing these estimated aftershock distributions to the distribution of observed aftershocks we obtain maps of the large earthquake slip. We verify that this mapping procedure correctly recover simulated fault slip, and matches independent estimates of slip for two California earthquakes. For the 2018 Mw 7.1 Anchorage Alaska earthquake we show that these slip maps, combined with precise aftershock relocations, allow determination of which fault the earthquake ruptured, and which secondary faults and geologic structures were activated in its energetic aftershock sequence.

# 1 Introduction

Earthquakes are well described by elastic rebound, the sudden release of elastic strain energy stored in a rock mass (Reid & Lawson 1908; Scholz 2019). This energy release occurs mainly at seismogenic depth, a brittle zone many kilometers below the Earth's surface (Sibson 1982; Marone & Scholz 1988), and is clearly manifested by slip across a fault. Comprehensive understanding of large earthquakes and their hazard requires mapping the orientation, extent, amplitude and time-variation of this fault slip. Such *finite-fault slip maps* are typically obtained from static or kinematic inversion of measured or calculated surface phenomena due to fault slip at depth. These surface observations include fault offsets, shaking intensity and time-varying or differential ground motion from seismogram, geodetic and space-based data such as global navigation satellite system (GNSS) and interferometric synthetic aperture radar (InSAR) (Olson & Apsel 1982; Hartzell & Heaton 1983; Ide 2007; Mai *et al.* 2016). The resulting finite-faulting maps depend on the selection, quality and distribution of observations, resolution of the inversion, and on numerous assumptions and parameters, such as a pre-defined, 2D fault surface geometry, models for earth structure, rupture speed and rise-time, an inversion method, and smoothing constraints or other regularization; consequently, finite-faulting maps for a single earthquake produced by different studies are often clearly different (Harris & Segall 1987; Beresnev 2003; Bos & Spakman 2003; Hartzell *et al.* 2007; Ide 2007; Minson *et al.* 2013; Mai *et al.* 2016; K. Wang *et al.* 2020).

The release of strain energy during earthquake slip also produces aftershocks and seismicity—phenomena occurring at seismogenic depth around an earthquake rupture. Aftershocks and seismicity form effective, in-situ observations that can provide information on fault slip. Earthquake slip releases strain energy and perturbs stress in the surrounding rock mass through stress transfer (Chinnery 1963). Under the Coulomb failure stress hypothesis, slip on a fault occurs when shear stress is sufficiently high and normal stress sufficiently low to overcome frictional resistance to sliding. In accordance with these concepts, stress changes due to an earthquake rupture will favor triggering of surrounding aftershocks within the positive lobes of the 3D spatial distribution of change in Coulomb failure stress (ΔCFS) (Stein & Lisowski 1983; Oppenheimer *et al.* 1988; King *et al.* 1994). Numerous studies show that, given a reasonable finite-faulting model for a mainshock, forward calculation of ΔCFS shows positive and negative lobes that often correlate well with observed distributions of occurrence or lack of aftershocks, respectively (Oppenheimer *et al.* 1988; King *et al.* 1994; Harris & Simpson 1996; Harris 1998; Shinji Toda *et al.* 1998; Stein 1999; Shinji Toda *et al.* 2011; Sato *et al.* 2012).

Here we introduce and illustrate a straightforward, imaging procedure for mapping finite-fault slip (or tensile opening) directly from seismicity and aftershocks. For specified source and receiver fault orientations, we construct 3D maps of extended source slip through correlation of point-source ΔCFS kernels across the spatial distribution of seismicity around a source earthquake. This *seismicity-stress* procedure finds, in accordance with the physics of the Coulomb failure stress criteria, a 3D distribution of relative, finite-fault slip compatible with the surrounding distribution of seismicity. Application of the procedure is mainly dependent on the availability of multi-scale precise relocations of seismicity surrounding a source zone of earthquake rupture, aseismic slip, fault creep, dyke intrusion or other strain source, and, for ΔCFS kernel calculation, on the specification of a shear dislocation or tensional mechanism for the source, a shear slip mechanism and faulting plane for receiver faults, and distance parameters for masking of the singularity at the ΔCFS point-source. Seismicity-stress slip maps might be used as prior constraint on fault geometry and slip for other slip inversion methods, for quasi data-driven aftershock forecasting, to aid in rapid shaking characterization and perhaps tsunami early-warning, to obtain finite-faulting information for large earthquakes occurring before the advent of modern seismic and geodetic instruments, and to identify possible regions of aseismic slip driving foreshock sequences.

We confirm that the seismicity-stress procedure correctly maps a synthetic slip distribution on two rectangular fault patches given random seismicity predicted by ΔCFS for the synthetic slip. We next use background seismicity and aftershocks along the Parkfield California segment of the San Andreas fault to produce seismicity-stress slip maps that match well other estimates of co- and post-seismic slip for the 2004 Mw 6.0 Parkfield earthquake, and of adjacent, long-term slip related to fast creep. We further confirm agreement between seismicity-stress slip maps and finite-faulting results from other methods for the 2021 Mw 6.0 Antelope Valley California normal-faulting earthquake. Finally, we show for the 2018 Mw 7.1 Anchorage Alaska, normal-faulting intra-slab earthquake how the seismicity-stress procedure, combined with multi-scale precise hypocenter relocation, resolves the enigma of which mainshock faulting plane ruptured (the gently east-dipping plane), and identifies seismotectonic structures in the slab activated in the energetic aftershock sequence.

## 2 The seismicity-stress procedure for mapping relative finite-fault slip in 3D

If an earthquake fault is considered a simple frictional surface with cohesion, then, under the Coulomb failure criteria (Coulomb 1773; Weiss *et al.* 2016), fault slip occurs when shear stress in the direction of slip becomes sufficiently high to overcome cohesion, and normal stress sufficiently low to remove frictional resistance to sliding. Though absolute stresses on a fault cannot be calculated, the change in stresses due to a source earthquake rupture with specified, finite-fault slip can be determined throughout an elastic volume (Okada 1992). For specified receiver fault orientations, these stress changes allows determination of the change in Coulomb failure stress, ΔCFS,

$$\Delta \text{CFS} = \Delta\tau + \mu'\Delta\sigma_n, \qquad (1)$$

where $\Delta\tau$ is change in shear stress on the fault, positive in the direction of slip, $\Delta\sigma_n$ is change in normal stress on the fault, positive for unclamping, and $\mu'$ an effective coefficient of friction which includes the effects of pore pressure changes (Stein & Lisowski 1983; Oppenheimer *et al.*

1988; King *et al.* 1994; Stein 1999). Stress changes due to the source slip will favor triggering of aftershocks or larger earthquakes on receiver faults within the positive lobes of the 3D spatial distribution of ΔCFS, and inhibit the occurrence of aftershocks or larger earthquakes in the negative lobes, often called stress shadows.

The 3D, scalar ΔCFS field due to finite-fault slip can be obtained through integration over the fault of a point-source, ΔCFS field derived from the stresses due to a point shear dislocation or tensile opening in an elastic half-space (Haskell 1964; Kikuchi & Kanamori 1982; Okada 1992; King *et al.* 1994; Materna & Wong 2023). This procedure is a forward calculation, with the resulting ΔCFS field often assessed by how well it agrees with the surrounding distribution of aftershocks and larger earthquakes. The corresponding inverse problem is to infer a distribution of finite-fault slip given the 3D distribution of aftershocks or other seismicity and the physics of the Coulomb failure stress criteria. Here we infer finite-fault slip for a target source event through an imaging methodology: for specified source and receiver fault orientations, we obtain 3D maps of source slip through correlation of point-source ΔCFS kernels across the spatial distribution of post-event seismicity around the target event. This *seismicity-stress* procedure finds a 3D distribution of relative finite-fault slip which can explain, in accordance with the physics of the Coulomb failure stress criteria, the surrounding distribution of post-event seismicity.

*Correlation of ΔCFS kernels over post-event seismicity*

The similarity of two signals or functions can be measured with cross-correlation. Here we are interested in the similarity of a point-source ΔCFS kernel field as it moves across an unknown, seismicity rate change field $\Delta S$ represented by a distribution of observed, post-source seismicity $\hat{S}$. To map this similarity in 3D we cross-correlate the two fields: as the point-source kernel origin is shifted across the seismicity in all three spatial dimensions, the dot product between the shifted kernel field and the seismicity field provides a measure of similarity which is assigned to the current point-source origin. Given a point-source ΔCFS kernel field and a seismicity distribution $\hat{S}$ in some time window, both sampled on a 3D grid with indices ($i, j, k$), a seismicity-stress, finite-fault slip field $F$ on the grid is obtained as,

$$F(i,j,k) = \sum_l \sum_m \sum_n \Delta CFS(l+i, m+j, n+k) \hat{S}(l,m,n), \qquad (2)$$

where the range of each summation index, *l, m, n*, is limited by the bounds of the *ΔCFS* and $\hat{S}$ field grids.

The use of a cross-correlation between a Green's function (the point-source ΔCFS kernel) and observations (the seismicity) to infer a model (fault slip) is well established as adjoint operator back-projection or imaging, which forms an approximate, but practical and robust alternative to formal inversion (Kawakatsu & Montagner 2008; Claerbout 2010; Fukahata *et al.* 2014). The approximation arises in part because imaging ignores physics-based, weighting or scaling functions between the model and data spaces, e.g. the normal matrix in least-squares matrix inversion is implicitly replaced by the identity matrix (Claerbout 1992). In consequence, the seismicity-stress field $F$ is an imaging or brightness function but not a formally defined physical quantity, and so here we refer to the field $F$ as a measure of *relative* fault dislocation or fracture opening.

A point-source ΔCFS kernel field is fully sampled across the study volume and typically forms a pattern of four positive and four negative lobes for a double-couple, shear dislocation source (Figure 2.1). In contrast, observed seismicity, $\hat{S}$, may or may not occur in areas of positive ΔCFS due to chance, and may occur in areas of negative ΔCFS at a reduced rate (Hardebeck & Harris 2022). Also, seismicity will be absent in parts of the surrounding rock mass that do not support brittle fracture or lack a sufficient density of existing receiver faults and fractures (J. Liu *et al.* 2003; Barchi *et al.* 2021). Thus, the post-source seismicity estimate, $\hat{S}$, of the Δ$S$ field can only provide a sparse, noisy, and irregular sampling of the positive regions the unknown, true ΔCFS field for the source event. In order to avoid bias due to this incomplete sampling of true Δ$S$ provided by the seismicity, we accumulate the spatial distribution of observed seismicity, $\hat{S}$, on a 3D grid using a value of +1 for cells containing seismicity and a value of 0 for other cells. Ideally, we would assign -1 to cells without seismicity and in the negative parts of the true ΔCFS, but this information is unavailable. We should also assign -1 to cells with inhibited or reduced seismicity rates if these can eventually be determined, for example through comparison of pre- and post-event seismicity.

Thus, under the physics of ΔCFS, Δ$S$ values of +1 flag areas where seismicity may be favored by the point-source dislocation or opening, while 0 values flag areas where the seismicity may be either favored or is suppressed. In cross-correlation between the gridded ΔCFS kernel and Δ$S$ seismicity fields, the aftershock grid +1 values can contribute to the correlation coefficient, while the 0 values will provide no information to and not change the coefficient. The resulting correlation coefficient should be somewhat insensitive to missing aftershocks in areas of true positive source event ΔCFS and to the impossibility of identifying lack of aftershocks in areas of true negative ΔCFS, as long as the distribution of available aftershocks in 3D approximates sufficient well the positive source event ΔCFS field.

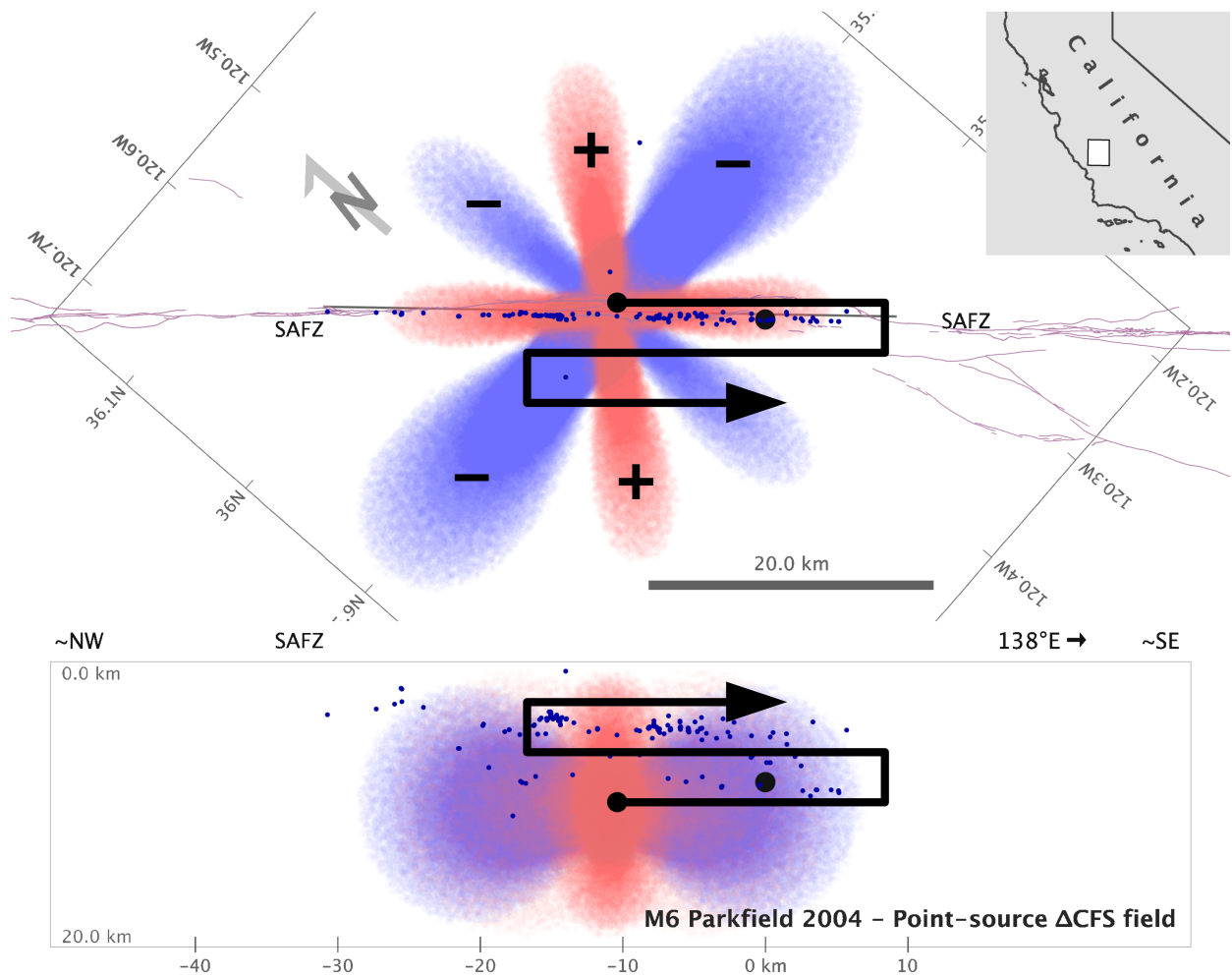

**Figure 2.1** Point-source ΔCFS kernel field for vertical, right-lateral strike slip source and receiver faults corresponding the geometry of the 2004 Parkfield sequence. The ΔCFS field is plotted as a 3D density cloud (red positive; blue negative) and is masked to avoid singularities in the stress calculations (zero mask radius $R_m$ = 1.0 km, decay length, $L_m$ = 10 km). The position and depth of the plotted kernel field in the study area is arbitrary. Blue dots show 4 hours of aftershocks after the 2004 Parkfield mainshock (large black dot). Arrow paths represents 3D shifting and correlation of the point-source kernel across the seismicity. Gray line shows strike of source fault, chosen to align with aftershocks and strike of the San Andreas fault zone (SAFZ). Purple lines show faults from the USGS Quaternary fault and fold database for the United States.

*Precision of relocated seismicity*

Meaningful application of the seismicity-stress procedure requires multi-scale precise relocations of seismicity surrounding a zone of earthquake slip or fault creep. That is, the distribution and relative positions of relocated hypocenters should represent well the true distribution of seismicity over the range of scales from the size of the study area down to the desired resolution of finite-faulting maps. Otherwise, errors in the hypocenter locations will map directly through correlation with the ΔCFS kernels into distortions or reduced resolution in the finite-faulting maps. Thus the relocated seismicity might use, at least, carefully determined station static corrections, or, preferably, explicitly multi-scale corrections such as source-specific, station traveltime corrections (SSST; Richards-Dinger & Shearer 2000).

Here we obtain multi-scale high-precision earthquake relocations with the NLL-SSST-coherence procedure (NLL-SC; Lomax & Savvaidis 2022; Lomax & Henry 2023). NLL-SC achieves multi-scale precises hypocenter location through the combination of SSST corrections and stacking of probabilistic locations for nearby event based on inter-event waveform coherence. NLL-SC is based on the NonLinLoc location algorithm (Lomax *et al.* 2000, 2014); NLL hereafter), which performs efficient, global sampling to obtain an estimate of the posterior probability density function in 3D space for hypocenter location. This probability density function provides a comprehensive description of likely hypocentral locations and their uncertainty, and enables application of the waveform coherence relocation. Within NLL, we use the equal differential-timing likelihood function (Zhou 1994; Font *et al.* 2004; Lomax 2005, 2008; Lomax *et al.* 2014), which is highly robust in the presence of outlier data caused by large error in phase identification, measured arrival-times or predicted travel-times, and thus helps stabilize the generation of SSST corrections which depend strongly on arrival residuals.

*The seismicity-stress procedure*

Given a target source event such as an earthquake, aseismic slip, fault creep or dyke opening, application of the seismicity-stress procedure to map finite-fault slip requires multi-scale precise relocations of post-event seismicity in a study volume surrounding a source event, and specification of a fault mechanism for the source and fault orientation and slip parameters for the receiver seismicity. The procedure includes the following steps:

- Specify a source mechanism (e.g. double-couple shear dislocation or tensile fracture opening); for a point double-couple the causative source-slip plane does not need to be specified since the produced stress field is independent of which of the two planes slips. Specify a receiver fault plane orientation and slip; the receiver slip plane must be specified so that the shear and normal stresses produced by the source can be resolved onto the receiving fault plane.
Specify the elastic parameters shear modulus and Lamé's first parameter.

- Establish a point-source ΔCFS field for the specified source and receiver faults within a 3D grid larger than the study volume. Here, this ΔCFS kernel grid is assembled from the point-source ΔCFS fields for a stack of gridded, horizontal layers created using the Elastic_stresses tool (Materna & Wong 2023) for calculating elastic displacements, strains, stresses and ΔCFS. Appropriate elastic parameters are chosen for the average depth of seismicity and the point source is placed at sufficient depth to minimize effects of the free surface. Figure 1 shows the double-couple point-source ΔCFS field for a vertical, right-lateral strike slip sources and receivers corresponding to mechanisms of events along the Parkfield section of the San Andreas fault.

- Mask the ΔCFS kernel grid in a region around the point-source origin coordinates, $\mathbf{X}_{ps}$, to avoid locally very high values of ΔCFS due to singularities in the stress calculations (Okada 1992). This mask $M(X)$ is defined by,

$$M(X) = 0 \qquad\qquad\qquad\qquad\qquad \text{for } |\mathbf{X} - \mathbf{X}_{ps}| < R_m, \qquad (3)$$
$$M(X) = 1.0 - \exp[-(|\mathbf{X} - \mathbf{X}_{ps}| - R_m)^2 / L_m^2] \qquad \text{for } |\mathbf{X} - \mathbf{X}_{ps}| > R_m,$$

  where $\mathbf{X}$ is a grid location, $R_m$ is a radius around $\mathbf{X}_{ps}$ within which the mask is set to 0.0, and $L_m$ is a decay length controlling an inverse Gaussian rise of the mask value to 1.0 far from the point-source. Full masking of the singularity around the point-source is controlled by $R_m$, which may be typically set to 1-2 times the grid spacing. The decay length, $L_m$, sets the smoothness of the termination of this masking and controls the distance from the point-source to the peak positive and negative amplitudes in the kernel field. $L_m$ might typically be set to roughly the extent of denser, post-event seismicity around the target source event and around 10 times $R_m$.

- Accumulate the post-event seismicity into a similar size 3D grid by assigning +1 to a grid cell if there are one or more hypocenters in the cell and assigning 0 otherwise.

- Correlate the masked, ΔCFS kernel grid across the seismicity grid to produce a 3D grid of potential, relative, static finite-fault slip.

- Normalize the finite-fault grid to its maximum value. Here, to aid in visualization and analysis, we create a second, high-potential grid where values ≤ 0.5 are removed and remaining values rescaled so that values [0.5, 1.0] map linearly to [0.0, 1.0]; this grid shows the extent of faulting potential that is greater than half the peak value.

- Optionally, clip the finite-fault grids to the convex hull of post-event seismicity.

- Output the finite-fault slip grids for visualization and further analysis.

The resulting finite-fault grids maps a 3D distribution of static slip (or tensile opening) which explains, in accordance with the physics of ΔCFS, the surrounding distribution of post-event seismicity. Essentially, the procedure resolves complex, 3D distributions of finite-fault slip through reverse mapping of aftershock locations through the non-isotropic lobes of the ΔCFS field.

The seismicity-stress maps shows *relative* slip since no formal theory relating source faulting to the distribution or rate of seismicity is used and, moreover, the seismicity is simplified to counts of 0 or +1 on a grid. The maps show *potential* slip since they show the envelope of an ensemble of allowable slip solutions compatible with the seismicity, and not a single slip model that additionally satisfies physical constraints such as energy conservation and localization of slip to 2D surfaces. The timing of the slip is only constrained to have been before or during the time range of the seismicity; some of the nominally post-event seismicity may be a response to seismic or a-seismic slip before the target source event.

The seismicity-stress procedure is simple and rapid. On an 8-core desktop workstation, the calculation of a ΔCFS kernel grid takes around 1 minute. Subsequent cross-correlation with an aftershock grid and output of a 3D finite-fault grid takes around 10 seconds, mostly for input and output. Thus, after a large earthquake, given relocations of aftershocks and pre-calculated ΔCFS kernel grids for different source and receiver faulting parameters, sets of 3D finite-fault grids reflecting the evolving seismicity and different faulting scenarios can be generated in near real-time to aid in rapid hazard analysis and aftershock forecasting.

## 3  Validation for synthetic fault slip on a rectangular fault patches

To validate the seismicity-stress procedure in a controlled and simplified case we consider synthetic slip on a vertical right-lateral, strike-slip fault with strike ~N140°E, following the geometry of the Parkfield segment of the central San Andreas fault (SAF; Figure 2.1). Slip is imposed in two different sized patches on the fault, each of which has constant slip and the same total moment release (Figure 3.1). We generate a representative set of synthetic aftershocks that might be produced by the synthetic slip distribution by: 1) calculating the stress changes and ΔCFS field (without masking) for the synthetic slip using the Elastic_stresses tool (Materna & Wong 2023), 2) sampling in proportion to the amplitudes in the positive lobes of this field to obtain a set of 474 synthetic aftershocks.

We specify the same strike-slip faulting as the synthetic slip for source and receiver faults to construct a double-couple, point-source ΔCFS kernel field on a grid of 80 km x 80 km horizontal x 20 km in depth grid with cell size 0.5 km (Figure 2.1). We use a zero mask radius, $R_m$, of 1.0 km and decay length, $L_m$, of 10 km.

Correlation of the masked, ΔCFS kernel grid across the gridded, synthetic seismicity produces the 3D grid of potential, relative, static finite-fault slip represented in Figure 3.1. The position and lateral extent of the two patches of synthetic fault slip are well recovered within the high-potential areas of the finite-faulting field (red). However, these high-potential areas spread beyond the imposed slip in depth and perpendicular to the fault horizontally, likely due to the fairly dense seismicity within the positive lobes of the ΔCFS field for the synthetic slip in these directions. Apparently, the vertically extended, sheet-like lobes of the point-source ΔCFS kernel (Figure 2.1) correlate well with the seismicity when the kernel origin is displaced from the synthetic slip within some distance, likely related to the decay length, $L_m$, of the kernel masking.

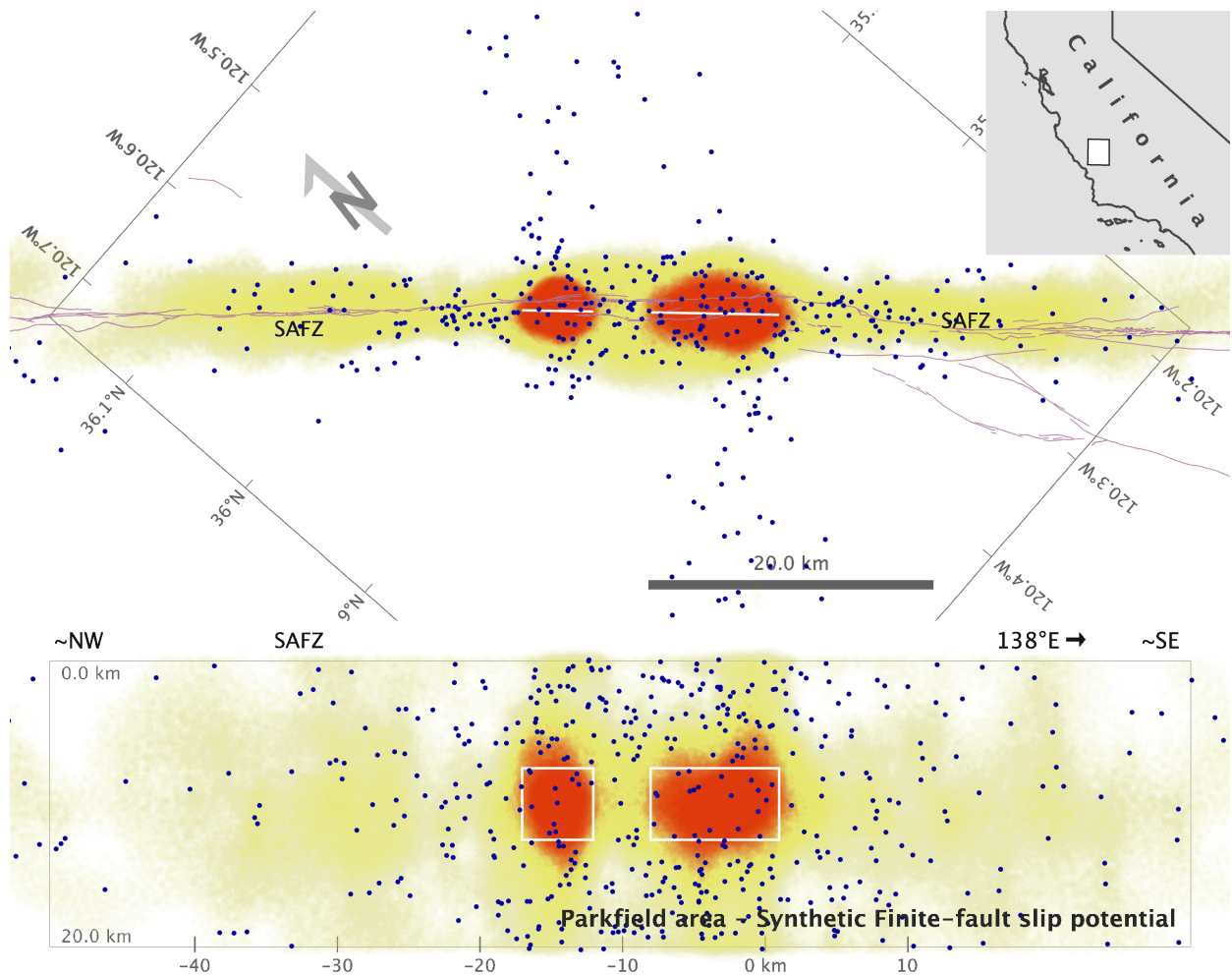

**Figure 3.1** Seismicity-stress, 3D finite-faulting potential inferred from seismicity (blue dots) sampled from the ΔCFS for synthetic rupture on a vertical fault (white rectangles). Synthetic slip for vertical, right-lateral strike-slip faulting corresponding to the geometry of the 2004 Parkfield sequence (Figure 2.1) is distributed uniformly within two patches: 5 x 5 km, to the northwest, and 5 x 10 km, to the southeast. Correlation is performed with a ΔCFS kernel for a point-source and receiver faults with the same faulting as the synthetic slip (Figure 2.1). The relative amplitude of the seismicity-stress finite-faulting field is shown as a 3D density cloud in tones of yellow for portions of the field with normalized correlation < 0.5 and in red for the high-potential portions of the field with normalized correlation ≥ 0.5. Field values are plotted in 3D for each grid cell as circles with radius increasing with correlation value; higher color saturation indicates higher correlation and/or deeper volumes of high correlation. The fields are not clipped to the convex hull of the seismicity. SAFZ indicates the San Andreas fault zone; purple lines show faults from the USGS Quaternary fault and fold database for the United States.

## 4 Tests on two California earthquake sequences

We test the seismicity-stress procedure through application to two recent earthquake sequences in California, the 2004, Mw 6.0 Parkfield sequence, a geometrically simple case of strike-slip faulting and most aftershocks on the same vertical plane, and the 2021, Mw 6.0 Antelope Valley sequence, with a normal faulting mainshock and a more complex, 3D distribution of aftershocks. In both cases we find generally good agreement with other finite-faulting results based on seismic waveform and geodetic data, but also a discrepancy in the location of highest slip relative to that found by inversion of seismic waveforms only.

*The 2004, Mw 6.0 Parkfield sequence*

We first consider the 2004 Mw 6.0 Parkfield, California sequence and background seismicity along the well instrumented, Parkfield segment of the central San Andreas fault (SAF; Figure 4.1). This segment is at the southeastern end of a 150 km long, near-linear stretch of the SAF which exhibits surface creep, repeating earthquakes and aseismic slip (Steinbrugge *et al.* 1960; Savage & Burford 1973; Nadeau & McEvilly 2004; Titus *et al.* 2006; Jolivet *et al.* 2015), and is just northwest of a long, locked section of the fault which last ruptured in the 1857 M 7.9 Fort Tejon earthquake (Bakun *et al.* 2005; Langbein *et al.* 2005). The 1966 M ~6 and 2004 Mw 6.0 Parkfield earthquakes both ruptured nearly the same fault patch on the Parkfield segment but initiated at opposite ends of this patch and propagated in opposing directions (Bakun *et al.* 2005). At seismogenic depth the Parkfield segment is considered to be a single, near-planar surface along the Southwest Fracture zone, which aligns linearly with the SAF to the northwest and southeast of Parkfield, but does not follow the curved, main San Andreas surface trace near Parkfield (Simpson *et al.* 2006; Thurber *et al.* 2006; Lomax & Henry 2023).

For seismicity-stress finite-faulting analysis, we consider here multi-scale precise, NLL-SC relocations for the Parkfield area (Lomax & Savvaidis 2022; Lomax & Henry 2023) using the earthquake catalog and phase arrival data provided by the Northern California Earthquake Data Center (NCEDC). These relocations places most background seismicity and 2004 aftershocks on a smooth, near-planar surface striking ~N140°E under and parallel to the Southwest Fracture zone. As the 2004 mainshock and most aftershocks and background seismicity have right-lateral strike-faults focal mechanisms (Thurber *et al.* 2006) with one fault plane in close agreement to the strike of the planar surface, we use vertical, strike-slip faulting with strike ~N140°E as the faulting mechanism for source and receiver faults. We construct a double-couple, point-source ΔCFS kernel field on a grid of 80 km x 80 km horizontal x 20 km in depth grid with cell size 0.5 km (Figure 2.1). We use a zero mask radius, $R_m$, of 1.0 km and decay length, $L_m$, of 10 km.

Correlation of the masked, ΔCFS kernel grid across the gridded NLL-SC seismicity for the first 4 hours after the 2004 Parkfield mainshock origin (at 2004-09-28 17:15:24 UTC) produces the 3D grid of potential, relative, static finite-fault slip represented in Figure 4.1. This finite-faulting field falls along and tightly around the vertical plane of seismicity and likely San Andreas fault plane at seismogenic depth, a correspondence that is constrained by the distribution of seismicity mainly along this plane and by the positive lobes of the ΔCFS kernel field aligned parallel to this plane (Figure 2.1).

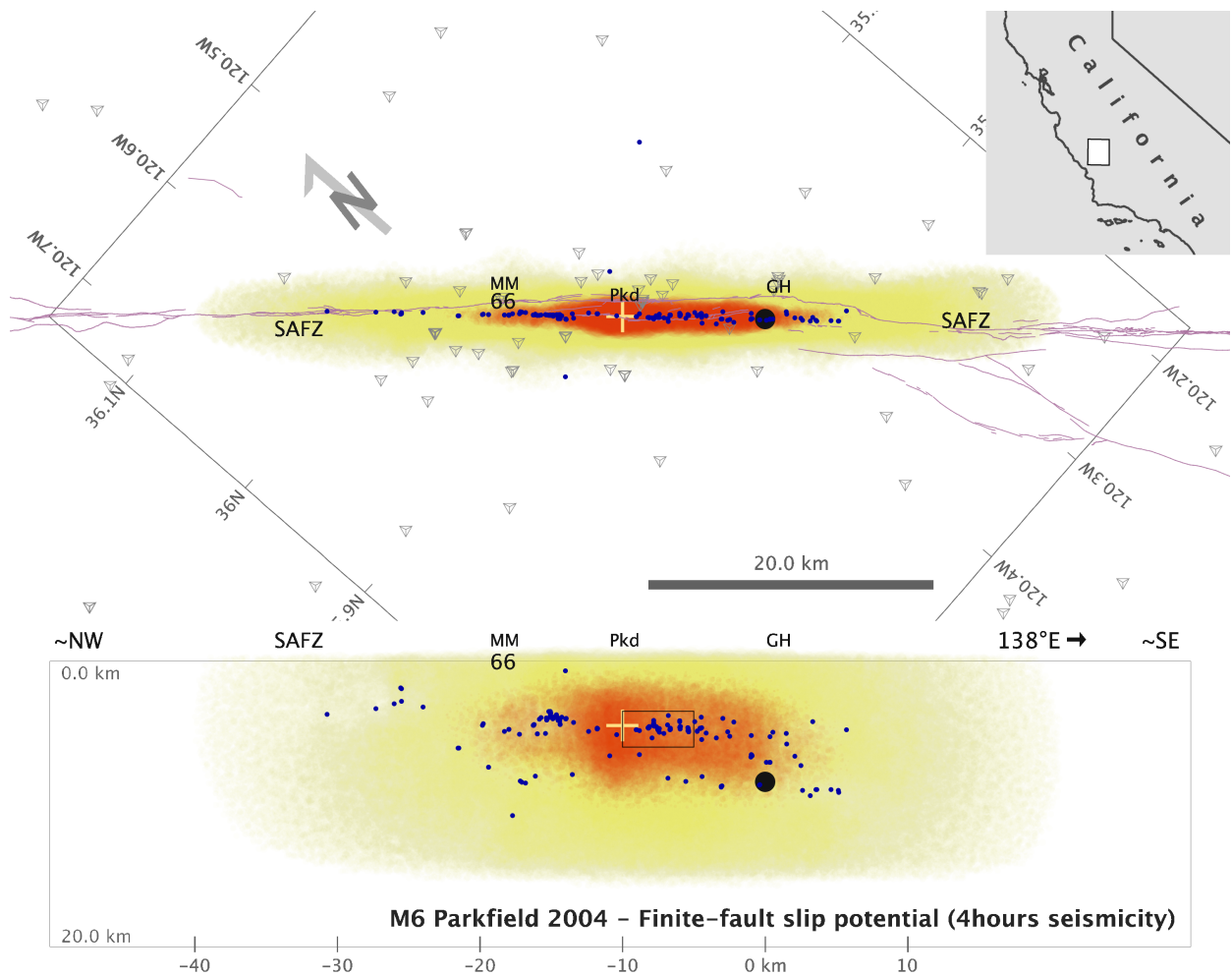

**Figure 4.1** Seismicity-stress, 3D finite-faulting potential inferred from the first 4 hours of aftershocks (blue dots) after the 2004 Parkfield mainshock (large black dot). Correlation is performed with a ΔCFS kernel for vertical, right-lateral strike-slip point-source and receiver faults (Figure 2.1). The relative amplitude of the seismicity-stress finite-faulting field is shown as a 3D density cloud in tones of yellow for portions of the field with normalized correlation < 0.5 and in red for the high-potential portions of the field with normalized correlation ≥ 0.5, the yellow cross shows the peak value in this field. Field values are plotted in 3D for each grid cell as circles with radius increasing with correlation value; higher color saturation indicates higher correlation and/or deeper volumes of high correlation. The fields are not clipped to the convex hull of the seismicity, but fading of the fields towards the limits of the plots may be an artifact of lack of available seismicity outside the plot and not absence of potential slip. The black box indicates clusters of seismicity that may be less active or silent before and after the 2004 sequence. The 1966 Parkfield mainshock epicenter (McEvilly *et al.* 1967) is indicated by "66", locations of seismographs shown as inverted triangles, other abbreviations are: SAFZ – San Andreas fault zone, MM – Middle Mountain, Pkd – Parkfield, GH – Gold Hill. Purple lines show faults from the USGS Quaternary fault and fold database for the United States.

The high-potential region of the finite-fault field (red) forms a horizontal band at ~3 – 7 km depths extending from just above to about 15 – 20 km northwest of the Mw 6.0 hypocenter, with the highest slip potential about 11 km northwest of the hypocenter in an ~4 km diameter, circular zone. The depth and along-fault ranges of this high potential slip region corresponds well to independent estimates of the geometry of co-seismic slip and energy radiation, especially given the diversity and uncertainty of these estimates (e.g. Hartzell *et al.* 2007). A main difference is that most slip inversions based on seismic and GNSS data find highest slip at around 15 – 20 km northwest of the Mw 6.0 hypocenter (Johanson *et al.* 2006; e.g. Langbein *et al.* 2006; P. Liu *et al.* 2006; L. Wang *et al.* 2012; Twardzik *et al.* 2012), though some inversions (e.g. Kim & Dreger 2008; Bruhat *et al.* 2011) find highest slip closer to the 11 km northwest of the hypocenter found by the seismicity-stress procedure. The position of the seismicity-stress peak agrees better with network and array based analyses of high-frequency seismic radiation from the 2004 mainshock which find the largest sources of high-frequency radiation ~7 – 13 km northwest of the hypocenter (Fletcher *et al.* 2006; Allmann & Shearer 2007; Fountoulakis & Evangelidis 2024).

The seismicity-stress procedure can also map temporal changes in areas of potential slip by application to seismicity in different time windows. Figure 4.2 shows 3D potential finite-fault slip maps along the SAF for seismicity in time windows before and after the 2004 Parkfield mainshock origin. The potential slip map for the first 4 hours after the mainshock (second row) shows the same results as Figure 4.1, with highest slip along the likely 2004 rupture zone and representing mainly co-seismic rupture. The potential slip maps for the 10 year periods 1984-1994 (Supplementary Figure S1) and 1994 to 2004 (top row in Figure 4.2) show highest slip potential along the southeastern end of the 150 km section of the central SAF where there is known background, fast creep, repeating earthquakes and aseismic slip, and the southeastern termination of this highest slip potential falls at the northwestern limit of the Parkfield locked asperity and 1966 and 2004 Parkfield rupture zone around Middle Mountain (Nadeau & McEvilly 2004; Titus *et al.* 2006; Maurer & Johnson 2014; Jolivet *et al.* 2015).

The potential slip map for the period from 1 week to 1 month after the 2004 mainshock shows highest slip potential along the 2004 co-seismic rupture zone (as defined by the potential slip map for the first 4 hours), but also further northwest into the fast-creeping section, which may represent capture of background seismicity and creep in this section, not slip triggered by the mainshock. This slip map also shows a broadening, deepening and shallowing of highest slip potential around the 2004 co-seismic rupture, which may represent after-slip of this rupture, in general agreement with previous studies (e.g. Johanson *et al.* 2006; Langbein *et al.* 2006; Freed 2007).

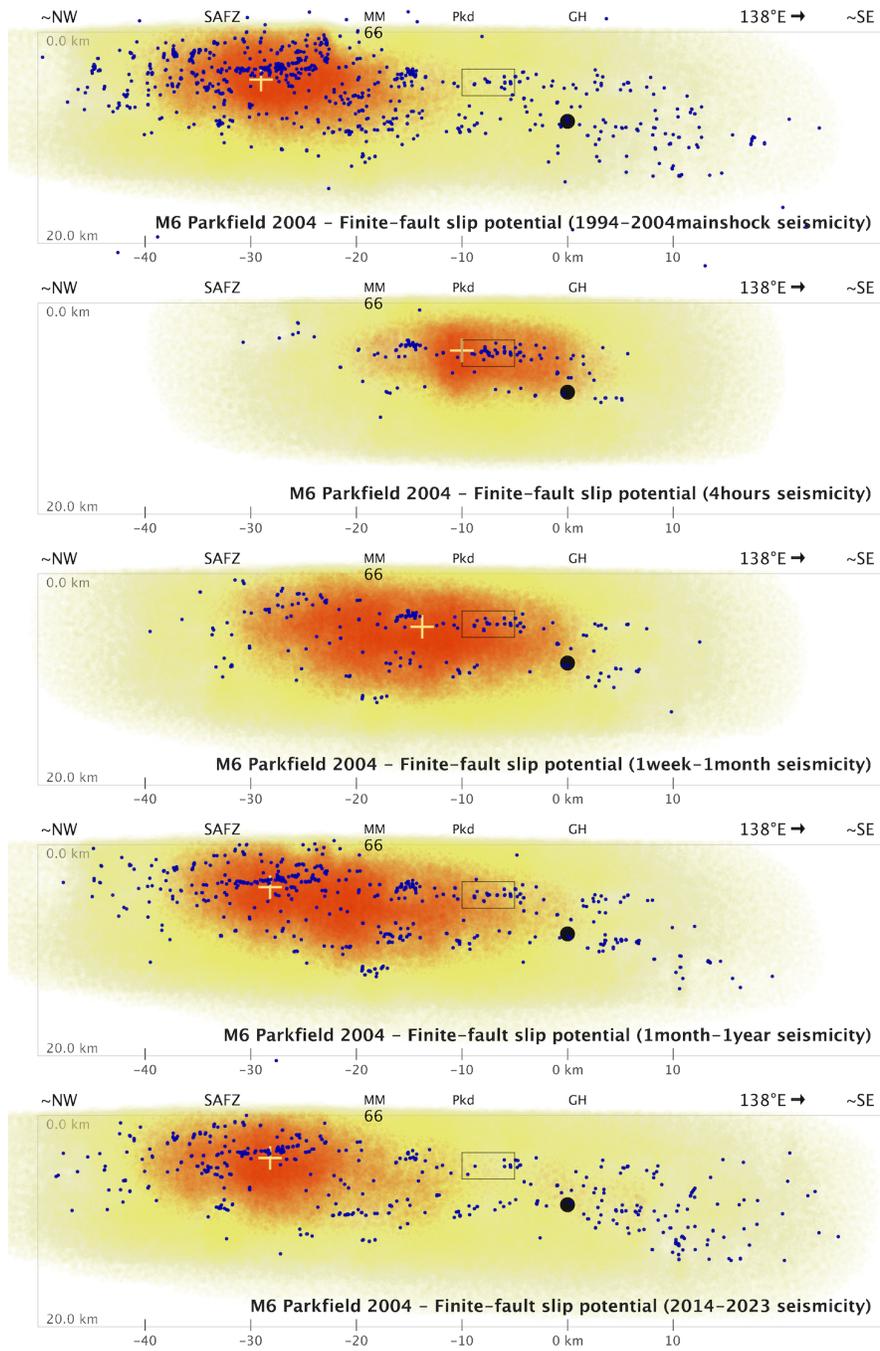

**Figure 4.2** Potential slip maps along the SAF for seismicity in a selection of time windows around the 2004 Parkfield mainshock. Maps labelled with date pairs show potential slip for aftershocks between those dates, those labelled with time spans show potential slip for that span after the 2004 mainshock origin. Map elements as in Figure 4.1. These and additional slip maps forming a contiguous set of time windows between 1984 and 2023 are presented in Supplementary Figure S1.

The potential slip map for the period from 1 month to 1 year after the 2004 mainshock shows a continuation of increasing response to background seismicity and creep to the northwest, and decreasing imaging of after-slip around the mainshock rupture. The peak of potential slip (yellow crosses in Figure 4.2) is shifted to the northwest into the fast-creeping section at about the same location as for the 10 year periods before and after the 2004 sequence. The slip map for the 9+ year period 2014 to 2023 (bottom row in Figure 4.2) is similar to those for 1984-1994 (Supplementary Figure S1) and 1994 to 2004 (top row in Figure 4.2), indicating that the majority of slip at depth and seismicity by 10 years after the 2004 mainshock is not distinguishable from the pre-mainshock background seismicity. Notably, the southeast limit of slip potential for the creeping section of the SAF as imaged by background seismicity from 1984-2004 and from 2014 to 2023 corresponds closely to the northwest limit of higher, 2004 co-seismic slip potential around 15 – 20 km northwest of the 2004 hypocenter (Figure 4.1).

The nature of the seismicity-stress methodology entails that these space-time variations in potential slip reflect local changes in density and extent of clusters of seismicity along the SAF in the Parkfield area. The overall distribution of clusters of seismicity, however, appears stable throughout the period 1994 to 2023, despite the occurrence of the Mw 6.0 2004 rupture, which supports the interpretation that seismically active and silent patches of the fault reflect differences in rheological and geometrical properties of the fault, and not transient stress variations (Thurber *et al.* 2006). However, there is indication that some clusters of seismicity were less active or silent after, and perhaps before, the 2004 sequence; for example, within the upper streak of events from about 5 to km northwest of the Mw 6.0 hypocenter (box in Figures 4.1, 4.2 and Supplementary Figure S1).

*The 2021, Mw 6.0 Antelope Valley sequence*

The 2021, Mw 6.0 Antelope Valley, California earthquake occurred in an area of normal fault-bounded basins within the complex, Walker Lane zone of trans-tensional deformation between the Sierra Nevada batholith to the west and the Basin and Range extensional province to the east (Wesnousky 2005; Wesnousky *et al.* 2012; Pierce *et al.* 2021). Estimates for the 2021 mainshock show hypocentral and centroid depths of ~8 – 10 km and an approximately north-south striking, normal-faulting mechanism with causative fault plane dipping ~45° to the east based on the overall geometry of aftershocks (USGS 2017; Pollitz *et al.* 2022; K. Wang *et al.* 2023).

We perform multi-scale precise, NLL-SC relocations (Lomax & Savvaidis 2022; Lomax & Henry 2023) for the 2021 Antelope Valley sequence using the earthquakes catalog and phase arrival data for the period 2021-01-01 to 2023-07-14 provided by NCEDC and the Nevada Seismological Laboratory (NSL). For seismicity-stress finite-faulting analysis of the 2021 Antelope Valley sequence we use the first 4 days of relocated aftershocks after the Mw 6.0 mainshock origin (at 2021-07-08 22:48:59 UTC). We use the NSL regional moment tensor (rCMT; Ichinose *et al.* 2003; available from USGS 2017) (nodal planes 1 / 2: strike = 194° / 356°, rake = 42° / 49°, dip = -76° / -102°) for source faulting and the east dipping plane of this mechanism for receiver faults to generate a point-source ΔCFS kernel (Supplementary Figure S2). We note here that first-motion mechanisms we obtain for our 2021 Antelope Valley relocations show a mix of normal faulting aftershocks compatible with the mainshock moment tensors and strike-slip aftershocks, where all focal mechanisms share an east-west orientation for the tensional axis. Thus future seismicity-stress analysis of this event should examine the case of using a point-source ΔCFS kernel with strike-slip receiver faults, and also application separately to normal and strike-slip faulting aftershock sets. In this study we examine two cases of receiver fault plane orientation in our analysis of the 2018 Mw 7.1 Anchorage Alaska sequence.

This seismicity-stress analysis produces the 3D grid of potential, relative, static finite-fault slip shown in Figure 4.3. The high-potential area of the finite-faulting field (red) forms an ~8 × 3 km, tabular patch at about 6 – 9 km depth with north-south orientation and dip parallel to the east-dipping fault plane of the rCMT and other centroid estimates (USGS 2017). The geometry of this high potential slip is constrained primarily by the distribution of seismicity around and mainly above the tabular patch along the east-dipping plane, and secondarily by the shallower seismicity to the east; most of this seismicity falls within positive lobes of the ΔCFS kernel fields (Supplementary Figure S2) for point sources centered within the tabular patch.

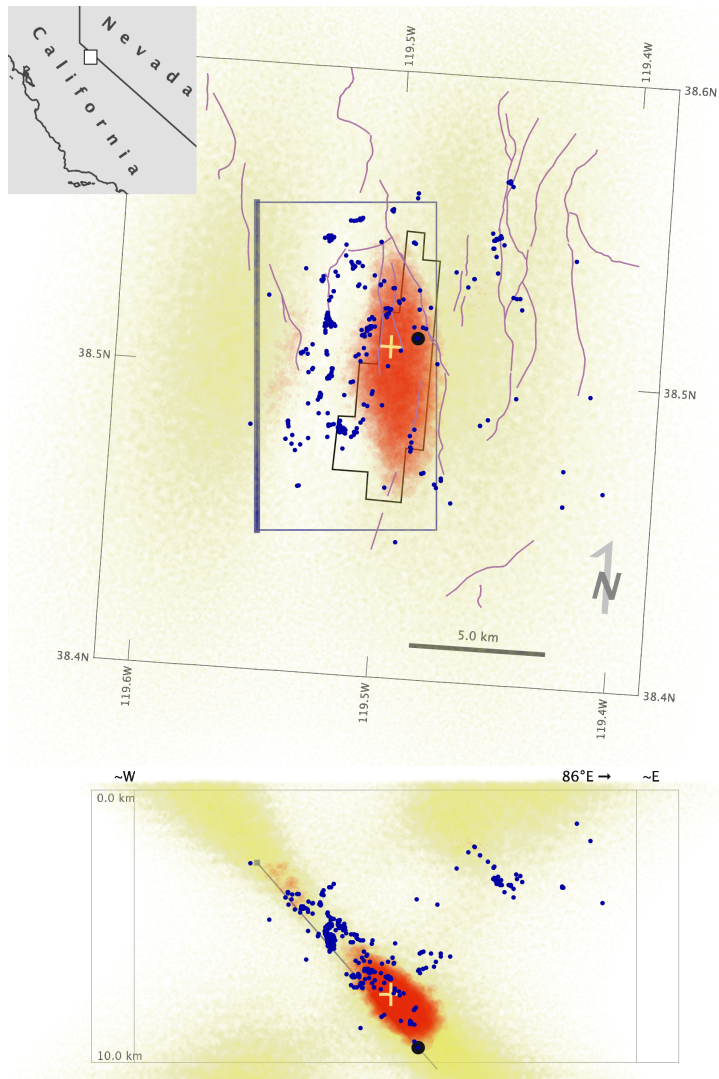

**Figure 4.3** Seismicity-stress, 3D finite-faulting potential slip maps inferred from the first 4 days of aftershocks (blue dots) after the 2021 Antelope Valley mainshock (large black dot). Map (upper) and profile (lower) views are rotated to the strike (4°W) of the NSL rCMT mainshock mechanism (gray rectangle with thick line on upper edge, positioned to intersect the relocated mainshock hypocenter and with arbitrary depth limits). Correlation is performed with a ΔCFS kernel (Supplementary Figure S2) for a normal-faulting point-source specified by the NSL rCMT and receiver faults corresponding to the east-dipping solution of the NSL rCMT. The relative amplitude of the seismicity-stress finite-faulting field is shown as a 3D density cloud in tones of yellow for portions of the field with normalized correlation < 0.5. The high-potential portions of the field with normalized correlation ≥ 0.5 is clipped to the convex hull of the seismicity and shown as a red density cloud, the yellow cross shows the peak value in this field. Field values are plotted in 3D for each grid cell as circles with radius increasing with correlation value; higher color saturation indicates higher correlation and/or deeper volumes of high correlation. The black polygon outlines higher-slip areas (> ~20 cm) of the preferred, kinematic slip model of Wang et al. (2023) based on geodetic and seismic data, shifted ~1 km ENE to align their mainshock epicenter with that found here (large black dot). Purple lines show faults from the USGS Quaternary fault and fold database for the United States

The dipping, 3D tabular patch of high-potential, seismicity-stress slip is compatible with rupture on a 2D surface parallel to the east-dipping plane of the for the rCMT and other centroid estimates for mainshock. However, the vertical position of the patch suggest a rupture plane shifted about 1 km shallower than a similarly oriented plane passing through the relocated hypocenter. This shift could be real, suggesting the hypocenter is not on the main slip plane, or a result of bias in the mainshock hypocenter depth relative to aftershocks depths due difficulty in picking arrivals, especially S, on larger amplitude and longer period large earthquake waveforms (Lomax 2020). Recall that the seismicity-stress procedure does not require prior specification of the mainshock hypocenter location or a mainshock fault plane depth or orientation, though the specifications of source and receiver faults put constraints on the recovered orientation of volumes of higher mainshock slip.

The high-potential, tabular patch of seismicity-stress finite-faulting matches well the position, extent and orientation of the USGS finite-fault model for the 2021 Antelope Valley mainshock based on teleseismic and regional seismic waveforms (USGS 2017), but only shows a general match in overall position to a more complex slip model found by Pollitz et al. (2022) using seismic and geodetic data. The seismicity-stress high-potential patch also matches well the extent and general orientation of higher slip area for the finite-source model of Wang et al. (2023) from joint, kinematic inversion of seismic waveforms, InSAR, and GNSS data (black polygon in Figure 4.3). One notable difference between these results is that the peak potential slip for the seismicity-stress model (yellow cross in Figure 4.3) is about 2 km north of the peak of slip in the Wang et al. (2023) model, the latter being further from the hypocenter, similar to our results for peak slip for the 2004 Parkfield mainshock. However, for two models based only on geodetic data shown by Wang et al. (2023, their figure 9) the area of peak slip overlaps with that of the seismicity-stress maps.

## 5 Application to the 2018, Mw 7.1 Anchorage, Alaska intra-slab earthquake sequence

The 2018, Mw 7.1 Anchorage, Alaska earthquake (Ruppert & Witter 2019) occurred within the north-northwest subducting Pacific plate under southern Alaska at an intermediate depth of about 45 km (Figure 5.1). The earthquake was felt throughout the greater Anchorage area, where it generated the strongest shaking since the great, 1964, M 9.2 Alaska megathrust earthquake, and caused moderate but widespread damage (West *et al.* 2019). Several point-source moment tensors for the 2018 mainshock (USGS 2017) show an approximately north-south striking, normal-faulting mechanism with one fault plane gently dipping ~30° to the east and the other steeply dipping ~60° to the west, consistent with down-dip extension in the subducting slab. The earthquake produced a highly productive aftershock sequence with the majority of moment tensors for larger aftershocks showing mechanisms similar to the mainshock (Ruppert *et al.* 2019; West *et al.* 2019; Drolet *et al.* 2022).

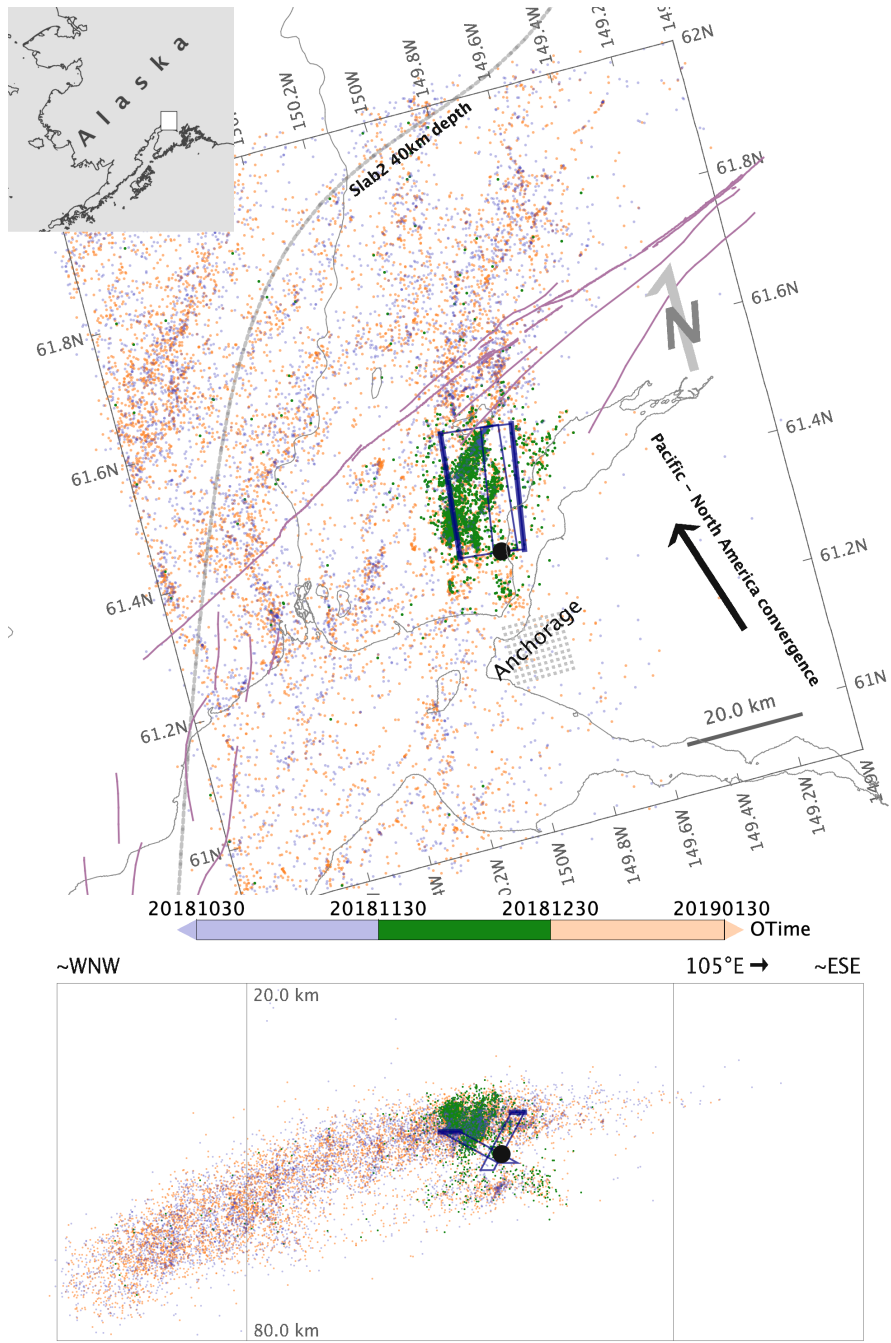

**Figure 5.1** Overview map of the 2018, Mw 7.1 Anchorage, Alaska sequence and background seismicity 2014-2022. Map (upper) and profile (lower) views are rotated to 15°E to roughly match the strike of the plane of dipping slab seismicity and corresponding strike of the Slab2 (Hayes 2018) 40km depth contour (thick gray line) to the west of the 2018 sequence. Note that the top of the profile view is at 20 km depth. NLL-SC relocated seismicity shown for: 2014 – 2018 mainshock (light blue), 2018 mainshock – 1 month after mainshock (green), 1 month after mainshock through 2022 (light orange); large black dot indicates the Mw 7.1 mainshock hypocenter. Arrow shows Pacific – North America relative plate convergence direction (Haeussler 2008); purple lines show surface faults from the USGS Quaternary fault and fold database for the United States.

Aftershocks and finite-faulting models of the 2018 mainshock give ambiguous constraint on which of the fault planes defined by point-source moment tensors hosted mainshock rupture. Some finite-fault models show a weak (USGS 2017; C. Liu *et al.* 2019) or strong (He *et al.* 2020) preference for mainshock slip on the steeply west-dipping plane, while it is also suggested that the rupture was simultaneous and conjugate on both planes (Guo *et al.* 2020). Thus, identification of the causative rupture plane for this earthquake remains an enigma, which we show can be resolved with the combination of multi-scale precise earthquake relocation and seismicity-stress finite-faulting analysis.

We perform multi-scale precise, NLL-SC relocations (Lomax & Savvaidis 2022; Lomax & Henry 2023) for seismicity in the area of the 2018 Anchorage sequence for the period 2014-01-01 to 2023-11-25 using catalog events and phase arrival data provided by the Alaska Earthquake Center. The hypocenters for aftershocks of the 2018, Mw 7.1 event (Figure 5.2) form two, main, northern and souther sub-clusters (Ruppert *et al.* 2019), as well as surrounding lineations and diffuse patches. The relocated seismicity since 2014 in the greater region around the 2018 sequence (Figure 5.1, Supplementary Movie S1) shows diffuse but organized seismicity mainly in an ~10 – 20 km thick, downward-bending tabular zone within the subducting Pacific plate around and below the general depth of the 2018 sequence. In profile view this seismicity shows a double seismic zone (Hasegawa *et al.* 1978; Brudzinski *et al.* 2007) in the descending Pacific slab (Ratchkovsky *et al.* 1997) under the area of the main 2018 sequence. In map view there is a larger scale organization and moderate clustering of seismicity around north to northeast trending lineations, though these do not show a simple relation to the direction of Pacific – North American plate convergence or to the strike of the dipping slab (see 40 km depth contour of Slab2 in Figure 5.2). However, along a northwest-southeast trend passing near the 2018 sequence, these lineations exhibit possible changes in orientations and character that could be related to the proximity of the southwest boundary in the subducting slab of the Yakutat microplate (Haeussler 2008), and to a change in strike of the dipping slab apparent in the 40 km depth contour of Slab2. Multi-scale precise relocation of seismicity over a larger area is necessary to verify and better interpret these trends and features in the seismicity.

For seismicity-stress finite-faulting analysis of the 2018 sequence, we use the first day of relocated aftershocks after the Mw 7.1 mainshock origin (at 2018-11-30 17:29:29 UTC). We construct point-source ΔCFS kernels using, for source faulting, the focal-mechanism of the USGS W-phase mainshock moment-tensor (USGS-WCMT; USGS 2017), (nodal planes 1 / 2:

strike = 6° / 189°, rake = 28° / 189°, dip = -93° / -88°), which is representative of other point-source moment tensor solutions for the mainshock.

For the receiver faults we considered two cases: the gently east- and steeply west-dipping fault planes of the USGS-WCMT mechanism. Only the case of west-dipping receiver faults (ΔCFS kernel shown in Supplementary Figure S3) produces a seismicity-stress map (Figure 5.2, Supplementary Movie S2) with a main patch of highest potential slip (red) that is a good candidate for mainshock rupture: this main patch 1) is tabular, and sub-parallel to and contains one of the USGS-WCMT fault planes through the hypocenter (the east-dipping plane), 2) abuts the mainshock hypocenter, and 3) contains few aftershock hypocenters but is bounded by aftershocks, especially far from the hypocenter and in the hanging wall, extensional quadrant of rupture.

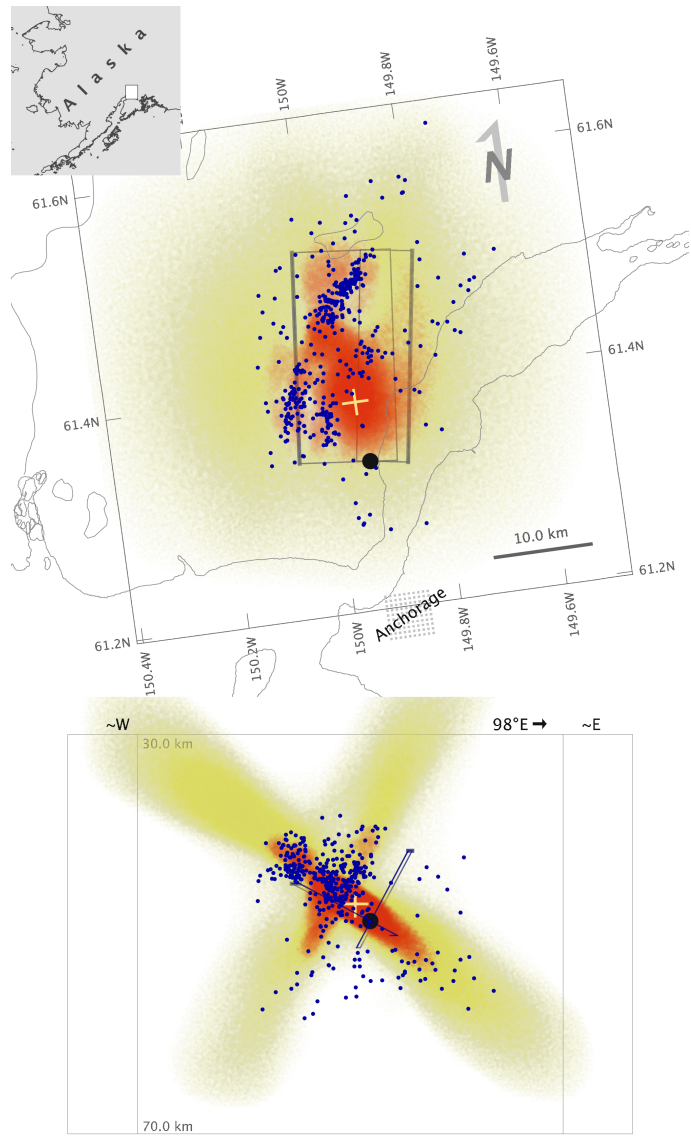

**Figure 5.2** Seismicity-stress, 3D finite-faulting potential slip maps for west-dipping receiver faults inferred

from the first 1 day of aftershocks (blue dots) after the 2018 Mw 7.1 Anchorage, Alaska mainshock (large black dot). Map (upper) and profile (lower) views are rotated to the strike of the neutral axis (8°E) of the USGS-WCMT mainshock mechanism (gray rectangle with thick line on upper edge, positioned to intersect the relocated mainshock hypocenter and with arbitrary upper and lower depth limits). Correlation is performed with a ΔCFS kernel for the normal-faulting, USGS-WCMT mechanism point-source and for receiver faults corresponding to the west-dipping fault of the USGS-WCMT. The relative amplitude of the seismicity-stress finite-faulting field is shown as a 3D density cloud in tones of yellow for portions of the field with normalized correlation < 0.5. The high-potential portions of the field with normalized correlation ≥ 0.5 is clipped to the convex hull of the seismicity and shown as a red density cloud, the yellow cross shows the peak value in this field. Field values are plotted in 3D for each grid cell as circles with radius increasing with correlation value; higher color saturation indicates higher correlation and/or deeper volumes of high correlation.

In contrast, the seismicity-stress map for east-dipping receiver faults (Supplementary Figure S4) has a main patch of highest potential slip in a steeply west-dipping, tabular region that is sub-parallel to the west-dipping USGS-WCMT fault plane, but offset about 5 km to the west, is further than 10 km from the mainshock hypocenter, and contains numerous aftershocks, thus not a good candidate for mainshock rupture. This seismicity-stress map does contain a secondary, tabular patch of potential slip that is parallel to the east-dipping USGS-WCMT plane, but this patch is weaker than and floats much higher above the east-dipping fault plane than the main patch obtained with west-dipping receiver faults. For neither receiver fault orientations do the seismicity-stress maps show larger potential slip in a patch overlapping and parallel to the west-dipping USGS-WCMT plane, which is weakly to strongly preferred for mainshock rupture in most previous studies (USGS 2017; C. Liu *et al.* 2019; Guo *et al.* 2020; He *et al.* 2020).

Thus the seismicity-stress analysis strongly suggests a mainshock faulting geometry with slip on the gently east-dipping fault plane, and supports rupture towards the north with peak-slip about 5 km north of the hypocenter, and a main rupture area (slip potential greater than half the peak value) of about 10 x 10 km (Figure 5.2). This rupture geometry matches well the non-preferred, joint seismic-geodetic finite-fault inversion of Liu et al. (2019; their supporting information figure S10) for the east-dipping fault model, except that the extent of their main rupture patch (slip greater than half the peak value) is larger at around 20 x 20 km and their main slip is further from the hypocenter at about 10 km than obtained with the seismicity-stress analysis.

# 6  Discussion

We have introduced a seismicity-stress imaging procedure for 3D mapping of earthquake finite-fault slip (or tensile opening) directly from seismicity and aftershocks. For specified source and receiver faulting, the procedure generate seismicity-stress maps through correlation of point-source Coulomb failure stress change (ΔCFS) kernels with the seismicity around an area of earthquake or other seismic slip. These maps show potential, relative, static fault slip as the envelope of an ensemble of slip solutions compatible with the surrounding seismicity given the physics of the Coulomb failure stress criteria. The reverse mapping of aftershock locations through the non-isotropic lobes of the ΔCFS field allows the method to resolve complex, 3D distributions of finite-fault slip, and makes it different from basic, geometrical interpretation of faulting from aftershock distributions (e.g. Mendoza & Hartzell 1988).

Application of the seismicity-stress procedure requires multi-scale precise relocations of seismicity surrounding a source zone of earthquake rupture, aseismic slip, fault creep, dyke intrusion or other strain source. Calculation of the point-source, ΔCFS kernels requires specification of a shear dislocation or tensional mechanism for the source, a shear slip mechanism and faulting plane for receiver faults, and distance parameters for masking of the singularity at the ΔCFS point-source.

The seismicity-stress method is based on information from seismicity at the depth of and surrounding the area of target seismic slip, which makes it qualitatively different from finite-fault inversion procedures which use surface observations. With the dense seismic sensor coverage available in many areas of significant seismicity and earthquake hazard, and precise earthquake location procedures such as SSST (Richards-Dinger & Shearer 2000) and NLL-SC (Lomax & Savvaidis 2022; Lomax & Henry 2023), the relative positions of seismic events across multiple scales can be determined with much smaller error (e.g. 100's of meters) than the typical scale length of interest and resolution in finite-fault mapping (e.g multi kilometer). Thus, with aftershock or background seismicity that fills an adequate volume around the target slip area, the seismicity-stress analysis can give strong, 3D constraint on potential slip.

Besides using information at depth, the seismicity-stress procedure has additional important differences with other methods for determining finite faulting. The seismicity-stress procedure does not require definition of the position and orientation of a fixed surface as locus for fault slip or fracture opening, nor does it require or make use of a hypocentral location or other initiation point for target slip; instead the procedure maps potential slip throughout the 3D volume covered by the seismicity. This unrestricted 3D mapping provides more freedom for identifying the location and orientation of main slip, as we saw with the case of resolving ambiguous rupture orientation for the 2018 Anchorage earthquake. The procedure also does not require definition of parameters like rupture speed and rise-time, and, as an imaging method, does not involve a formal inversion with its dependence on smoothing constraints or other regularization. However, the seismicity-stress procedure does not directly quantify directions or magnitude of slip and moment release, nor does it produce temporal information on rupture evolution. Given information such as a point-source, centroid moment tensor and elastic properties for a source rupture, it is possible that mean and perhaps variations in slip, stress-drop and other quantities can be derived from the area of high seismicity-stress slip potential. The seismicity-stress procedure can be used to map migration of areas of slip, after-slip and creep over longer time periods, as we illustrated for the Parkfield stretch of the SAF

Though we have only explored shear dislocation sources in this study, the seismicity-stress methodology should be valid for mapping extended sources of tensional fault or fracture opening, such as volcanic dyke intrusions, as long as these sources produce seismicity on surrounding faults or fractures. The theory for the change in stresses due to a tensile source is well developed (Okada 1992), and existing algorithms for calculating strain and ΔCFS support these sources (e.g., Shingi Toda *et al.* 2011; Materna & Wong 2023).

*Further applications and uses of seismicity-stress slip maps*

There are a number of potentially significant applications of seismicity-stress analysis and uses for the obtained slip maps. We outline a few of these applications and uses here, without attempting to explore further their implementation or performance.

Seismicity-stress slip maps can provide prior constraint on the number and geometry of faults and orientation of slip for finite-fault inversions that use seismic, geodetic and other surface measures. This procedure would combine the strong, 3D spatial information on slip derived by seismicity-stress from the distribution of seismicity at depth around a rupture with the temporal and more explicitly quantitative information derived by formal inversion methods that use surface data.

Another important use for seismicity-stress maps could be estimation of the spatial likelihood of future seismicity as a component of aftershock forecasting (Steacy *et al.* 2004). This would be a quasi data-driven and physics based procedure: Aftershocks up to the current time are used to develop a seismicity-stress map, which is an extended distribution of weighted, ΔCFS point-sources. A weighted sum over the ΔCFS from each of these point-sources then provides an estimate of the full ΔCFS field for the extended mainshock rupture. This field represents the relative spatial likelihood of future aftershock occurrence, including for areas where aftershocks have not yet occurred.

Seismicity-stress slip maps may also aid in rapid shaking characterization and perhaps tsunami early-warning for large earthquakes when data transmission and basic event processing (e.g. determination of a mainshock mechanism or moment-tensor and relocation of early aftershocks) are available in near-realtime. In this case, the seismicity-stress slip information could be integrated into algorithms that estimate shaking intensity or model tsunami generation and propagation from finite faulting models. Depending on instrumental coverage and the aftershock productivity of a seismic sequence, seismicity-stress slip maps might be available within a few hours of the main event, as illustrated in the Parkfield application in this study.

A particularly consequential application of seismicity-stress analysis would be to obtain finite-faulting information for large earthquakes that occurred before the advent of modern seismic, geodetic and space-based instruments. Any early-instrumental event for which a faulting mechanism can be established and for which there are reliably located aftershocks in and around the source regions is a candidate for such analysis. For the largest events, aftershocks located with epicentral uncertainty much less than the extent of the expected rupture zone, and with little depth constraint or depth constrained by independent information (e.g. known megathrust depth profiles, recent seismicity, or regional tectonics) might qualify as reliably located.

Lastly, seismicity-stress analysis could be applied to foreshock sequences to identify possible regions of aseismic slip or creep that trigger the foreshocks through Coulomb stress transfer, and thus aid in distinguishing between cascading and pre-slip models of earthquake nucleation (Ellsworth & Beroza 1995; Mignan 2015). Seismicity-stress mapping using foreshock seismicity might resolve concentrated areas of aseismic slip active before a mainshock in a manner analogous to the seismicity-stress mapping of creep related slip and after-slip over longer time periods for the Parkfield application in this study. A lack of identification of concentrated pre-slip might provide supporting evidence for a cascading process for a foreshock sequence.

*Validation of the seismicity-stress procedure for mapping 3D finite-fault slip*

We verified in Section 3 that the seismicity-stress procedure correctly recovers the location and form of two patches of synthetic, double-couple slip on a rectangular fault, given a random sample of aftershocks distributed according to the theoretical ΔCFS field for the extended slip source (Figure 3.1). For clarity, this test uses a somewhat ideal distribution of seismicity that statistically follows the complete ΔCFS field. In practice, a heterogeneous distribution of ambient stresses, of existing faults and fractures, and of geologic units with varying mineralogy, heat flow, fabric, rheology and other properties will also affect strongly the distribution of seismicity (Collettini *et al.* 2009; Hardebeck 2022; Hardebeck & Harris 2022).

For a more realistic test of the seismicity-stress procedure we computed slip potential maps for the 2004 Mw 6.0 Parkfield CA and 2021 Mw 6.0 Antelope Valley CA earthquakes and compared them with published finite-fault inversions and other independent information on these events. For Parkfield the seismicity-stress analysis using vertical, strike-slip source and receiver faults and 4 hours of aftershocks recovers highest slip potential concentrated along a vertical plane, as expected for this stretch of the SAF (Figure 4.1). For Antelope Valley the analysis using 4 days of aftershocks and normal faulting source and receiver faults recovers concentrated slip potential around a dipping plane that matches the causative rupture plane expected from aftershocks and mainshock centroid moment-tensor solutions (Figure 4.3). Both of theses analyses produce maps of highest slip potential that match well the distribution of larger slip obtained from finite-fault inversions, though the seismicity-stress results place peak slip ~5 km closer to the mainshock hypocenter for Parkfield and ~2 km closer to the hypocenter for Antelope Valley. We discus this discrepancy in more detail below.

For Parkfield we also examined seismicity-stress maps for a number of time windows of up to two decades before and after the 2004 mainshock (Figure 4.3 and Supplementary Figure S1) These maps show possible migration of after-slip around the co-seismic rupture area in the weeks and months after the mainshock. They also show highest slip potential in the decades before the mainshock falls as expected in the well documented, fast-creeping stretch of the SAF just northwest of the 2004 and 1966 Parkfield rupture zone, and a slow recovery towards a similar slip distribution over the two decades after the 2004 mainshock. This temporal analysis illustrates use of the seismicity-stress procedure for longer term mapping of source areas of main co-seismic, post-seismic, creep related, and perhaps other types of slip.

*Discrepancy in location of highest-slip patch*

For the 2004, Mw 6.0 Parkfield sequence the seismicity-stress procedure maps a co-seismic finite-faulting field (Figure 4.1) that closely follows the vertical SAF, with areas of high potential slip that match well the distribution of main slip obtained by other methods, given their uncertainties. The patch of highest seismicity-stress slip potential locates about 11 km northwest of the 2004 hypocenter, which agrees with the location of principal sources from inversions of high-frequency seismic radiation, but not with most seismic and GNSS based slip inversions which find highest slip 5 km or more further to the northwest.

This discrepancy in position of maximum slip could be due to shortcomings in the seismicity-stress procedure or available aftershocks locations. However, seismicity-stress slip maps for background seismicity in the period 1984-2004 and from 2014 to 2023 (Figure 4.2 and Supplementary Figure S1) show a southeast limit zone of slip potential for the creeping section of the SAF around 15 – 20 km northwest of the 2004 hypocenter which is coincident with the northwest limit of the patch of highest, 2004 co-seismic slip potential. This abutting of slip regions suggests little or no slip deficit would be available for significant 2004 rupture as far as 20 km northwest of the hypocenter. In this case, the discrepancy in location of highest-slip patch could be due to the lack of GNSS sites to the southeast of the rupture zone (Houlié *et al.* 2014), or a bias in peak slip location in seismic waveform inversions.

For the 2021, Mw 6.0 Antelope Valley mainshock we found a similar, though smaller, discrepancy in peak slip location, with the seismicity-stress peak located ~2 km closer to the hypocenter than found through joint inversion of seismic waveforms and geodetic data (Wang et al. 2023). However, in this case the location of the seismicity-stress peak corresponds well to that of inversion by Wang et al. (2023) with geodetic data only.

A shift in peak slip away from the hypocenter with broadband seismic waveform inversion that is not present with high-frequency seismic inversion or geodetic inversion (nor with seismicity-stress mapping) suggests a bias unique to broadband seismic waveform inversion. Perhaps such slip inversion is biased towards imaging sources of high-amplitude, broadband stopping phases related to rupture arrest and rupture approaching the free surface (Savage 1965; Madariaga 1977; Madariaga & Ruiz 2016), and not towards imaging areas of strongest total moment release, which, if smooth, may radiate mainly low amplitude, long period waves (Madariaga 1977) which might not be observable. Similarly, if the true rupture is crack-like with a long rise time and strong, late slip around the middle of the fault (Madariaga 1977), or involves reversal of the direction of the slip front (e.g. Hicks *et al.* 2020), then inversions for which these cases are not allowed in the modeled rupture evolution might produce erroneous mapping of slip too far from the hypocenter. There could also be shifts in peak slip location due to directivity or other near-source waveform effects (Archuleta & Hartzell 1981) if they, for example, produce augmented, high-amplitude signal late on the waveforms for seismic sensors located above the fault or in the direction of rupture which are incorrectly mapped to high-amplitude slip near these sensors.

*The 2018, Mw 7.1 Anchorage, Alaska intra-slab earthquake sequence*

We generated seismicity-stress maps for the 2018, Mw 7.1 Anchorage sequence using the representative, USGS-WCMT mainshock mechanism to define the Coulomb point-source faulting and each of the fault planes from this mechanism to define two cases of receiver fault orientation and slip. The resulting pair of potential slip maps and relocated seismicity (Figures 5.2 and Supplementary Figure S4) strongly support the case of mainshock rupture on a gently east-dipping plane with most aftershocks involving normal slip on planes parallel to the steeply west-dipping, USGS-WCMT plane. The seismicity-stress maps and relocated aftershock seismicity do not support mainshock rupture on a steeply west-dipping plane.

The potential slip maps are mainly constrained by concentrations of aftershock activity along the four, positive lobes of the point-source ΔCFS kernel (Figure 5.2 and Supplementary Figure S4, Supplementary Movies S1 and S2) as it is displaced along the resolved, high slip regions. This constraint includes numerous aftershocks in the hanging wall and extensional quadrant of rupture, and also two lobes of aftershock seismicity under sequence extending into the lower part of the double seismic zone (Figures 5.1 and 5.2). We can attribute the activation of this deeper aftershock seismicity to the two, down-going positive lobes of the Coulomb stress field from extended mainshock rupture.

Our relocated seismicity shows many aftershocks in northern and southern sub-clusters on or near an ~20km long, ~ 8 km vertical, steeply northwest-dipping plane which strikes about 40°E (Figure 6.1, Supplementary Movies S1 and S2), markedly clockwise to the ~8°E strike of the USGS-WCMT planes. Seismicity along this plane is also apparent within the lineations in the pre- and post- sequence background seismicity (Figures 5.1 and 6.1). Other nearby aftershocks fall on shorter, oblique or conjugate trends on the east side of this 40°E striking plane, the ensemble suggesting a larger scale, perhaps rhomboidal fracture system. Drolet et al. (2022, their figure 6) identify some events near the 40°E striking plane with a moment-tensor fault plane sub-parallel to the plane, but abundant seismicity with slip on this plane seem incompatible with the seismicity-stress result that most aftershock slip should occur on normal faults sub-parallel to the west-dipping USGS-WCMT plane. This discrepancy may be related to the nature of brittle fracture in slabs at intermediate depth, perhaps involving volume processes such as dehydration embrittlement or thermal shear instability (Raleigh & Paterson 1965; Hobbs & Ord 1988; Kirby *et al.* 1996; Hasegawa & Nakajima 2017). For example, background and aftershock activity may occur on smaller scales as a complex volume-filling process in which individual slip planes are oblique to larger scale trends of seismicity, in analogy to crustal processes such as wrench tectonics (Anderson 1905; Sibson *et al.* 2012) or fault-fracture meshes (Sibson 1996). In this case the 40°E trend of aftershocks might not define a surface of active faulting, but instead a western limit to a zone of brittle failure, perhaps composed of fractures, faults and anomalous rheology formed during earlier plate bending at the outer rise or plate formation at the spreading ridge.

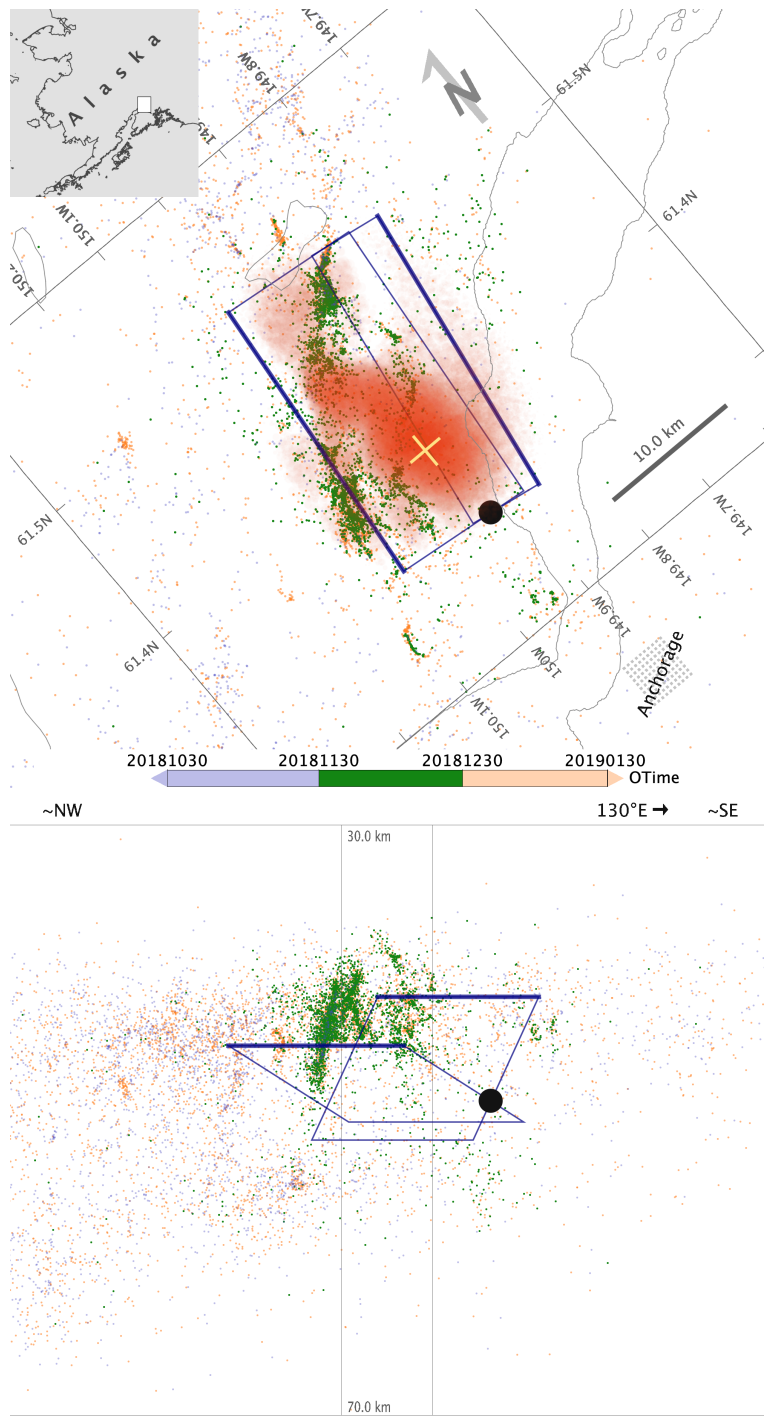

**Figure 6.1** Detail map of the 2018 Mw 7.1 Anchorage, Alaska sequence and background seismicity 2014-2022. Map (upper) and profile (lower) views are rotated to align along the steeply northwest dipping plane of aftershocks striking 40°E. Note that the top of the profile view is at 30 km depth. NLL-SC relocated seismicity shown for: 2014 to 2018 mainshock (light blue), 2018 mainshock to 1 month after mainshock (green), 1 month after mainshock through 2022 (light orange); large black dot indicates the Mw 7.1 mainshock hypocenter. The high-potential portions of the preferred seismicity-stress finite-faulting field from Figure 5.2 is shown in red in the map view, the yellow cross shows the peak value in this field.

The seismicity along and east of the 40°E striking plane appears rooted on the east-dipping, mainshock rupture plane, up-dip from the main slip and extending up to ~7 km above this plane (Figures 5.2 and 6.1). Thus much of the aftershock activity occurs off the main rupture plane in complex faulting in the hanging wall, and mainshock rupture (red field in Figure 6.1) occurs below a mainly aseismic gap between zones of aftershock seismicity, as previously proposed by Ruppert et al. (2019).

## 7 Conclusions

We have introduce a seismicity-stress imaging procedure for 3D mapping of finite-fault slip or tensile opening directly from seismicity and aftershocks following the physics of the Coulomb failure stress criteria. As this seismicity occurs at depth around an earthquake rupture or other seismic source, it provides strong constraint on the 3D distribution of potential, static fault slip. The seismicity-stress procedure is fairly simple and requires few assumptions; it does not require specification of the location and orientation of a 2D surface as locus for slip. Seismicity-stress maps may be useful as prior constraint for other slip inversion procedures, for quasi data-driven, physics based aftershock forecasting, and for rapid shaking characterization and perhaps tsunami early-warning. The seismicity-stress procedure should be applicable to map finite-faulting for early instrumental earthquakes, to search for aseismic slip or creep during foreshock sequences, and to other problems.

We verify that seismicity-stress maps correctly recover two patches of synthetic fault slip on a vertical fault, and match well the results from other inversion for finite-fault slip for the 2004 Mw 6.0 Parkfield CA and 2021 Mw 6.0 Antelope Valley CA earthquakes. We find indication of a discrepancy between locations of peak slip found with the seismicity-stress procedure and seismic and GNSS based slip inversions. It is not clear if this discrepancy reflects a limitation of the seismicity-stress procedure or of other slip inversion procedures.

Our seismicity-stress analysis of the 2018, Mw 7.1 Anchorage sequence shows normal-faulting mainshock slip on a gently south-east dipping plane, resolving the ambiguity in causative fault plane given by other finite-faulting analyses. The seismicity-stress map and multi-scale precise hypocenter relocations for the 2018 sequence show abundant aftershock activity in the hanging wall and extensional quadrant of mainshock rupture. Most aftershocks concentrate in a volume along and east of a 40°E striking plane, suggesting a rhomboidal fabric perhaps controlled by earlier plate bending and faulting. The analysis indicates two diffuse lobes of aftershocks descending into the lower part of a double, slab seismic zone are likely due to Coulomb stress triggering from mainshock rupture.

**Acknowledgments**

I gratefully thank to Pierre Henry and Kathryn Materna for discussion and comments that greatly improved this work, and to Jean Paul Ampuero for pointing out the relation between cross-correlation and adjoint operator imaging. Special thanks to the field crews, analysts, technicians and scientists who enabled the high-quality NCEDC, NSL and Alaska Earthquake Center arrival-time and event catalogs which made the precision of this work possible. LibreOffice (https://www.libreoffice.org) is used for word processing, spreadsheet calculations and drawings, and Zotero (https://www.zotero.org) for citation management. This work was supported by the author's personal funds.

**Data and code availability**

The supplementary information for this article includes Figure S1-4, and, in a repository (https://zenodo.org/doi/10.5281/zenodo.10939981), Movies S1 and S2 showing animated views of relocated seismicity and seismicity-stress results for the 2018 Mw 7.1 Anchorage, Alaska earthquake sequence presented in the main paper, Datasets S1-? containing CSV format catalogs of NLL-SSST-coherence relocations used in this study, and File S1 containing run scripts and related set-up, configuration and other meta-data files for seismicity-stress cases presented in this paper.

Earthquake catalogs and corresponding phase arrival times, waveforms and metadata were accessed through the Northern California Earthquake Data Center (NCEDC), http://doi.org/10.7932/NCEDC; information produced by the Alaska Earthquake Center was retrieved from the ANSS Comprehensive Catalog operated by the U.S. Geological Survey (USGS 2017). Fault trace data are from https://www.usgs.gov/programs/earthquake-hazards/faults.

Software and examples for running seismicity-stress mapping are available at: https://github.com/alomax/seismicity_stress

Earthquake relocations were performed with NonLinLoc (Lomax *et al.* 2001, 2014); http://www.alomax.net/nlloc, https://github.com/alomax/NonLinLoc) following the workflow presented in (Lomax & Savvaidis 2021) and using configuration metadata available in File S1. Stress calculations were performed with Elastic_stresses_py (Materna & Wong 2023), https://github.com/kmaterna/Elastic_stresses_py). SeismicityViewer (http://www.alomax.net/software) was used for 3D seismicity analysis and plotting, ObsPy (Beyreuther *et al.* 2010; Krischer *et al.* 2015), (http://obspy.org) for reading and processing seismicity catalogs and for coherence calculations.

**Competing interests**

The author declares having no competing interests.

# Mapping finite-fault slip with spatial correlation between seismicity and point-source Coulomb failure stress change

Anthony Lomax*[1]

[1]ALomax Scientific, Mouans-Sartoux, France

*Corresponding author: anthony@alomax.net

## Supplementary Material

**This PDF file includes:**

Figure S1-4

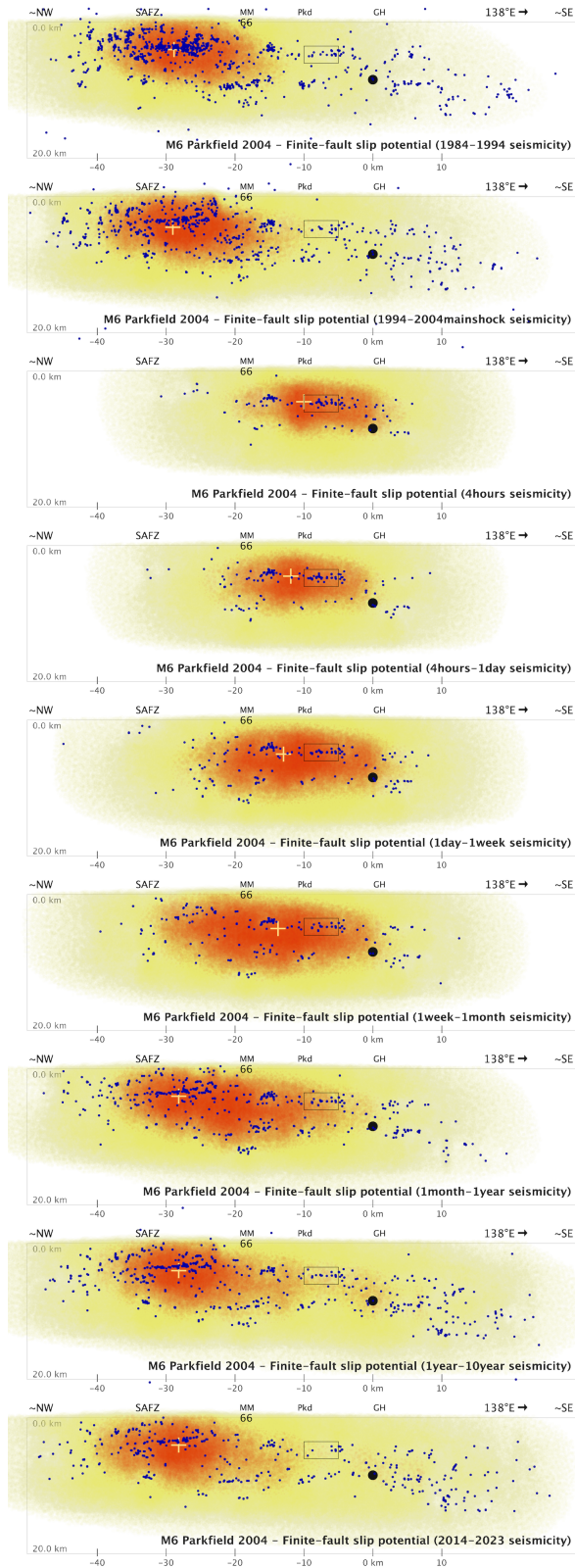

**Figure S1**. Potential slip maps along the SAF for seismicity in a contiguous set of time windows around the 2004 Parkfield sequence. Maps labelled with date pairs show potential slip for aftershocks (blue dots) between those dates, those labelled with time spans show potential slip for that span after the 2004

mainshock origin. The 2004 mainshock is indicated by the large black dot. Correlation is performed with a ΔCFS kernel for vertical, right-lateral strike slip point-source and receiver faults (Figure 2.1). The relative amplitude of the seismicity-stress finite-faulting field is shown as a 3D density cloud in tones of yellow for portions of the field with normalized correlation < 0.5 and in red for the high-potential portions of the field with normalized correlation ≥ 0.5. Field values are plotted in 3D for each grid cell as circles with radius increasing with correlation value; higher color saturation indicates higher correlation and/or deeper volumes of high correlation. The fields are not clipped to the convex hull of the seismicity, but fading of the fields towards the limits of the plots may be an artefact of lack of seismicity outside the plot and not absence of potential slip. The black box indicates clusters of seismicity that may be less active or silent before and after the 2004 sequence. The 1966 Parkfield mainshock epicenter (McEvilly et al. 1967) is indicated by "66", other abbreviations are: SAFZ – San Andreas fault zone, MM – Middle Mountain, Pkd – Parkfield, GH – Gold Hill.

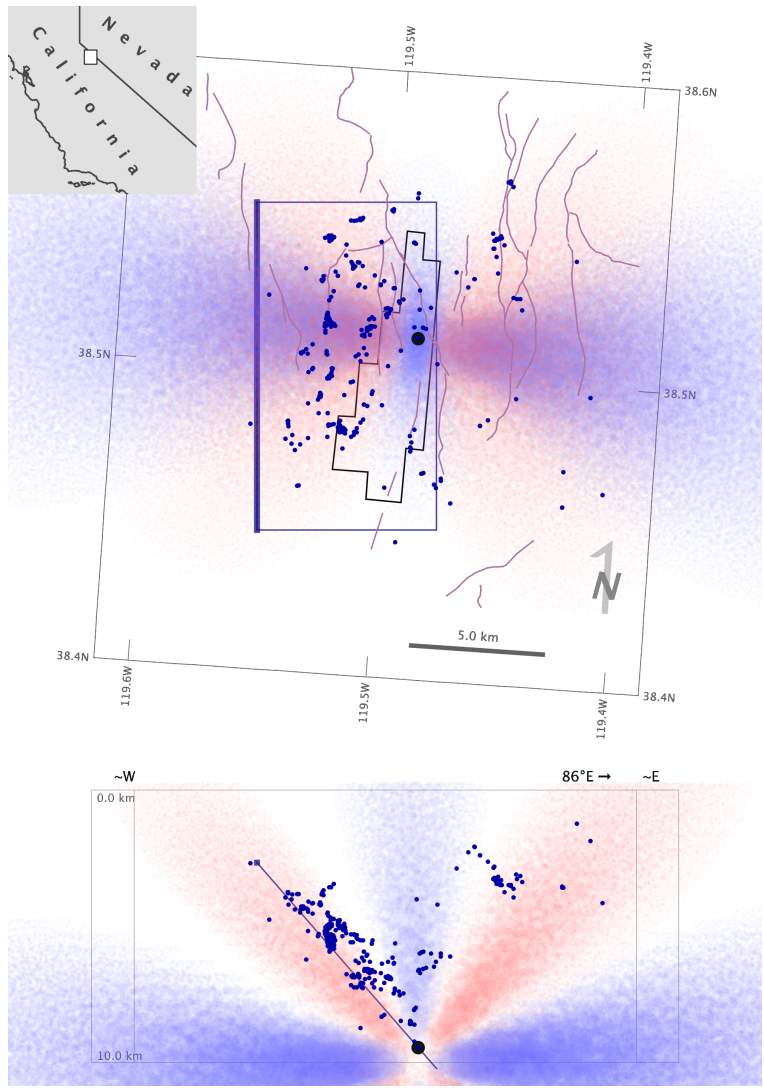

**Figure S2**. Point-source ΔCFS field (red positive; blue negative) for east-dipping, normal slip source and receiver faults corresponding the geometry of the 2021, Mw 6.0 Antelope Valley sequence. Blue dots show 4 days of aftershocks after the 2021 mainshock (large black dot). Gray rectangle with thick line on upper

edge shows orientation of east-dipping, NSL rCMT fault plane, positioned to intersect the relocated mainshock hypocenter and with arbitrary upper and lower depth limits. Purple lines show faults from the USGS Quaternary fault and fold database for the United States.

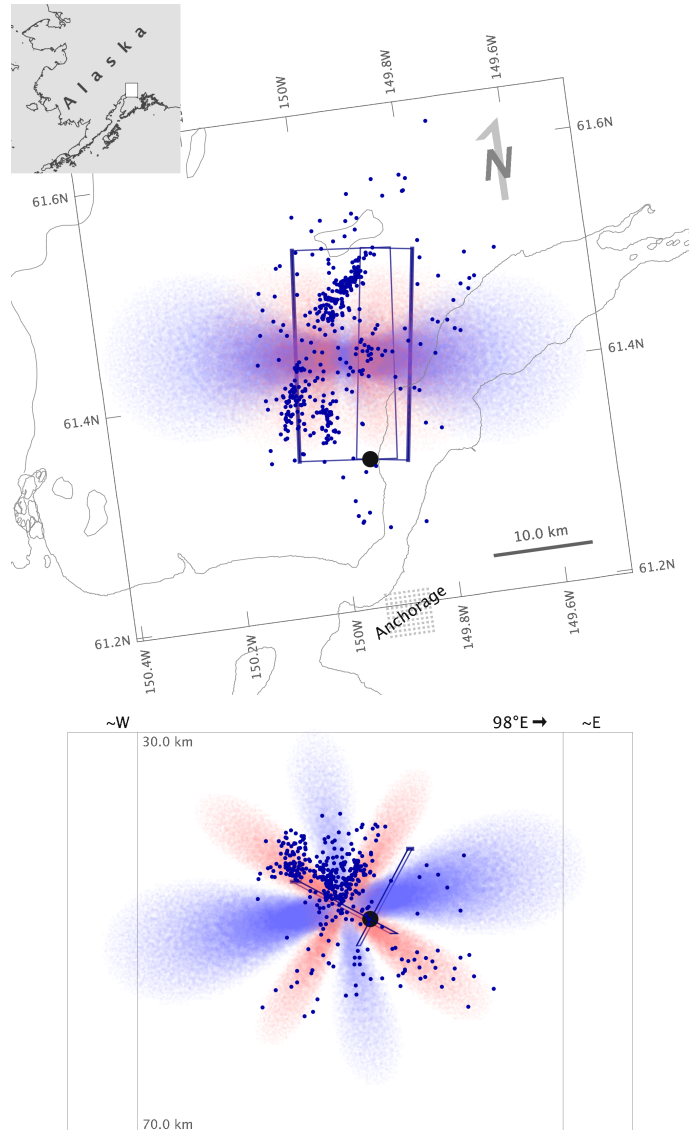

**Figure S3**. Point-source ΔCFS kernel field (red positive; blue negative) for the normal-faulting, USGS-WCMT mechanism as source and receiver faults corresponding to the west-dipping plane of the USGS-WCMT mechanism for the geometry of the 2018, Mw 7.1 Anchorage sequence. Blue dots show 1 day of aftershocks after the 2018 mainshock (large black dot). Gray rectangles with thick line on upper edge shows orientation of east- and west-dipping USGS-WCMT fault planes, positioned to intersect the relocated mainshock hypocenter and with arbitrary upper and lower depth limits.

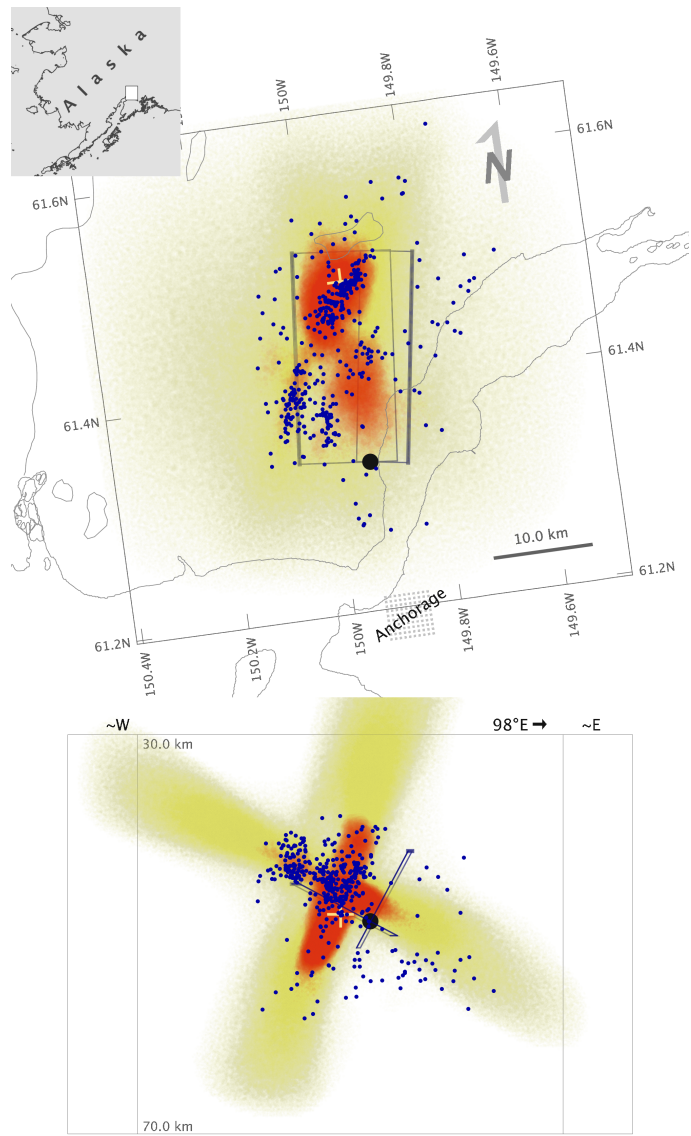

**Figure S4**. Seismicity-stress, 3D finite-faulting potential slip maps for east-dipping reciever faults inferred from the first 1 day of aftershocks (blue dots) after the 2018 Mw 7.1 Anchorage, Alaska mainshock (large black dot). Map (upper) and profile (lower) views are rotated to the strike of the neutral axis (8°E) of the USGS-WCMT mainshock mechanism (gray rectangles with thick line on upper edge, positioned to intersect the relocated mainshock hypocenter and with arbitrary upper and lower depth limits). Correlation is performed with a ΔCFS kernel for the normal-faulting, USGS-WCMT mechanism point-source and for receiver faults corresponding to the east-dipping fault of the USGS-WCMT. The relative amplitude of the seismicity-stress finite-faulting field is shown as a 3D density cloud in tones of yellow for portions of the field with normalized correlation < 0.5 and in red for the high-potential portions of the field with normalized correlation ≥ 0.5, the yellow cross shows the peak value in this field. Field values are plotted in 3D for each grid cell as circles with radius increasing with correlation value; higher color saturation indicates higher correlation and/or deeper volumes of high correlation.